\documentclass[fleqn,10pt]{wlscirep}
\usepackage[utf8]{inputenc}
\usepackage[T1]{fontenc}
\usepackage{multirow}
\title{West Australian Pandemic Response: The Black Swan of Black Swans - A Case Study}

\author[1]{David Cavanagh}
\author[2]{Mark Hoey}
\author[3]{Andrew Clark}
\author[4]{Michael Small}
\author[5]{Paul Bailey}
\author[6]{Jon Watson}

\affil[1]{Integrated Energy Pty Ltd, Como, Perth, Western Australia (email: david.cavanagh@integratedenergy.com.au)}
\affil[2]{Integrated Energy Pty Ltd, Como, Perth, Western Australia (email: mark.hoey@integratedenergy.com.au)}
\affil[3]{Integrated Energy Pty Ltd, Como, Perth, Western Australia (email:andrew.clark@integratedenergy.com.au)}
\affil [4]{Integrated Energy Pty Ltd, Como, Perth, Western Australia (email:michael.small@integratedenergy.com.au}
\affil [4]{Complex Systems Group, Department of Mathematics and Statistics, University of Western Australia, Crawley, Perth, Western Australia (email: michael.small@uwa.edu.au)}
\affil [4]{Mineral Resources, Commonwealth Scientific and Industrial Research Organisation, Kensington, Perth, Western Australia}
\affil [5]{Integrated Energy Pty Ltd, Como, Perth, Western Australia (email paul.bailey@integratedenergy.com.au)}
\affil [5]{St John Western Australia, Belmont, Western Australia (email paul.bailey@stjohnwa.com.au)}
\affil [6]{Faculty of Health and Medical Sciences, University of Western Australia, Crawley, Perth, WA (email:jon.watson@uwa.edu.au)}

\begin{abstract}
The COVID-19 Pandemic has been described as the global challenge of our time, an enormous human tragedy with dramatic economic impacts.  This paper describes the response and expected recovery process for Western Australia, where a rapid and effective response was implemented. This has enabled an early transition into an expected recovery both in health and economic terms.  The positive lessons learned from this experience are documented as they emerge in order to support other states and nations as they address this issue globally in the near-term and consider enduring improvements for the longer term.  While the authors have personal experience in the WA context, wider observations across Australia and selected international benchmarks are also included.  Key lessons include the importance of good health advice in Australia's interest; timely, synchronized and aligned action at all levels of government; a program of well communicated, aligned health and economic measures which support all in society allowing a very high level of appropriate community behaviour, ensuring the health system was not overloaded; innovation in telehealth, testing, pandemic modelling, and integrated operations which also allowed essential industries to continue; and strong border and travel controls with highly effective isolation preventing community spread, ultimately enabling rapid elimination of the disease from the hospital system.  In combination, these demonstrate that in the case of Western Australia the result of first eliminating the disease from the community, and then reopening the economy progressively at a strong pace, has enabled a world leading outcome in both in health and economic terms.  The lessons from this experience are widely applicable, shareable both as supporting service to other regions and through knowledge transfer.

\end{abstract}
\begin{document}

\flushbottom
\maketitle
\thispagestyle{empty}

\section*{Introduction}

The novel coronavirus SARS-CoV-2 (also known as COVID-19) that was first reported in December 2019, is now responsible for the first pandemic since 2009 H1N1 influenza. The world hasn’t experienced an event like this since the original H1N1 influenza pandemic, that of the 1918 Spanish flu. As of the 15th May 2020, COVID-19 had been reported in 187 countries, with more than 4.5 million cases and more than 300,000 confirmed global deaths\cite{RN95}. Although many countries have been able to slow the rate of spread of COVID-19, the number of cases globally continues to increase. The rapid spread of COVID-19 has led to a wide range of responses from different governments. These responses have included travel controls, both interstate and international, closures of schools and businesses, strict quarantine and isolation measures, implementation of testing regimes, and emergency funding of existing healthcare structures. These responses have been further complicated by worldwide shortages of personal protective equipment (PPE) and the burgeoning economic cost of these pandemic response measures. 

 Countries vary in terms of their inherent ability and preparedness to effectively navigate a pandemic. Population size, density and demographics, level of sanitation, quality and capacity of healthcare system, and the efficiency of government structures all affect the vulnerability of different countries. Combining these country specific factors with the varying country specific responses to the pandemic has seen a wide range of outcomes across the globe. 

The usual parameter with which to quantify the transmissibility of a pathogen is the basic reproductive number ($R_0$). It can be defined as the average number of secondary infections produced by a typical case of an infection in a population where everyone is susceptible\cite{RN97}.  The parameter $R_0$  must be greater than 1 for the number of cases to be increasing and therefore lead to an epidemic. There have been several estimates of the $R_0$  value for the COVID-19 pathogen with variation across different locations but a recent systematic review estimated it at 3.32 when uncontrolled\cite{RN96}. This number is determined by factors such as a person's duration of infectivity, the rate of contacts in the population, and the probability of infection being transmitted during contact. 

The rate of contacts in a population is a parameter that governments can change when bringing in quarantine, isolation and social distancing measures which reduce the $R_0$ value to the effective reproductive number $R_{eff}$. Reducing $R_{eff}$ slows the rate of increase of infected individuals and results in a lower peak of active cases. For countries aiming for pandemic suppression, the aim is to reduce $R_{eff}$ below 1 to reduce the number of cases to very low levels. For countries aiming for pandemic mitigation, the aim is to reduce $R_{eff}$ to a value above 1. This has the effect of ‘flattening the curve’ of pathogen transmission and reducing peak load on healthcare systems. A study recently analysed data for COVID-19 in 25 different countries and found that $R_{eff}$ consistently decreased after implementation of pandemic control measures\cite{RN98} . Australia in general, and Western Australia in particular, are examples of where $R_{eff}$ has been reduced below 1 for an extended period to the point where elimination of the pathogen, even if temporarily, is a possibility. The purpose of this paper is to chronicle the actions taken locally in Western Australia and nationally across Australia and summarise key learning points from the response to the pandemic. 

It is important to note that there is a significant delay between when a government control measure is implemented and when you would expect to see a change in the growth of new cases. The incubation period of a pathogen can be defined as the time between the initial exposure to the pathogen to when symptoms and signs are first apparent. Several studies have been performed estimating the median incubation period of COVID-19 with results ranging from 4 to 5.2 days\cite{RN102, RN101, RN100}. If we assume an extra 48-72 hour period is required for a symptomatic patient to be tested, return a positive test result and have that result recorded, there should be approximately a 7 day delay between a government control measure being implemented and the measure having an effect on the growth of new cases. 

\section*{Current pandemic status in Western Australia and nationally}

At the time of writing on 15th May 2020, Australia has successfully ‘flattened the curve’ on the COVID-19 pandemic. As of the 15th May 2020, Australia has had 7035 confirmed cases, 98 deaths and 6359 recoveries, leaving just 578 active cases across the country\cite{RN93}. Roughly 0.4\% or 27 new cases, were reported in the last 48 hours which contrasts to a peak of 830 cases in a 48 hour period over 27-28th March 2020\cite{RN94}. Between 20th April and the 15th May 2020, Australia has averaged approximately 15 new cases a day showing a two-week block of the pandemic well under control. 

In Western Australia as of the 15th May, the last confirmed case was on the 14th May 2020 and but there have only been 8 new cases since the 20th April 2020. This contrasts with the peak of the outbreak in Western Australia on the 30th March where 44 new cases were confirmed. As of the 15th May, 63,248 tests had been performed in Western Australia with 554 positive cases. Of these 554 positive cases, there have been 9 deaths and 538 recoveries, leaving 7 active cases in Western Australia. May 15th also marks the first day since the beginning of the outbreak that there are no active cases of COVID-19 in Perth hospitals\cite{RN99}. Most of the restrictions that were put in place to control the pandemic are still in place and discussion has now moved to how the state returns to a new normal.

\section*{The path through the pandemic}

Getting to Western Australia’s current position has not been easy. A wide range of measures have been implemented including international and state travel bans and the enforcement of strict social distancing measures. Overall, Australia and Western Australia have been successful in their pursuit of pandemic suppression. But this success has not been without its missteps in other states. These include the disembarkation of the Ruby Princess cruise-ship in Sydney, and local outbreaks in north-west Tasmania and in Melbourne. 

\subsection*{Timeline of events}

We divide the timeline of events into distinct periods as follows:

1.	26th January -- 16th March 2020: Time from the first case until a State of Emergency was declared in Western Australia. 

2.	16th March -- 20th April 2020: Time from State of Emergency declaration in Western Australia until the first 0 case day.

3.	20th April -- 15th May 2020: First 0 case day until first day with no active cases in Perth hospitals.

\subsection*{26th January -- 16th March 2020: Time from the first case until a State of Emergency was declared in Western Australia}
\begin{figure}[!ht]
\centering
\includegraphics[width=0.9\textwidth]{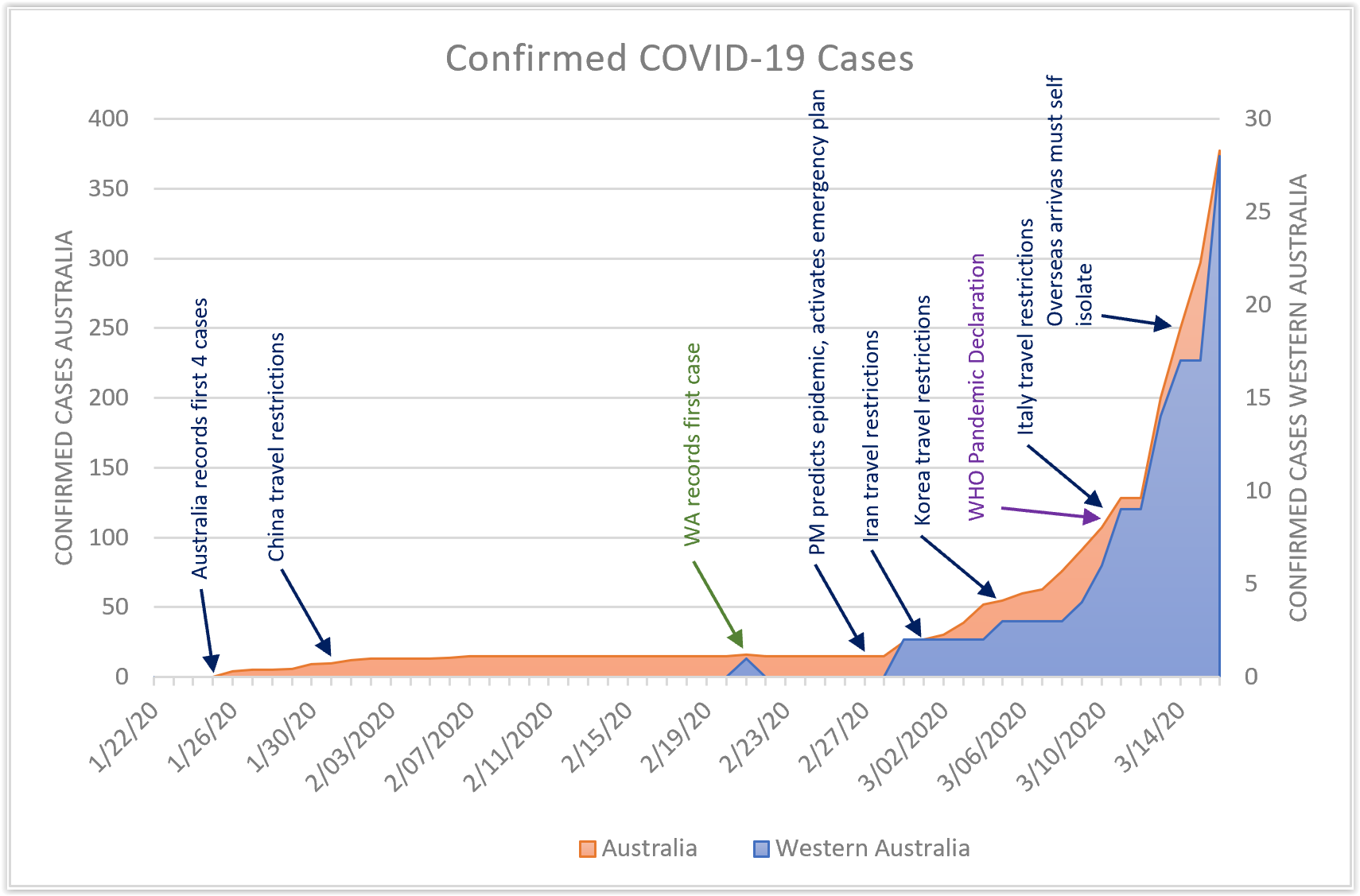}
\caption{Confirmed COVID-19 Cases from first case until West Australia State of Emergency Declaration}
\label{fig:Figure 1.}
\end{figure}

Fig. \ref{fig:Figure 1.} shows the early stages of the pandemic. In this time the Chief Health Officer had activated the existing Infectious Disease Emergency Management Plan, WA Health System (2017), and State Hazard Plan Human Biosecurity, WA Emergency Management Committee (2019), with a specific WA COVID Pandemic Response Plan then published and made publicly available by the Premier of WA on his Facebook page.  By the 16th March when a State of Emergency was declared in Western Australia, there were 377 confirmed cases in Australia and 28 confirmed cases in Western Australia. 

\begin{figure}[!ht]
\centering
\includegraphics[width=0.9\textwidth]{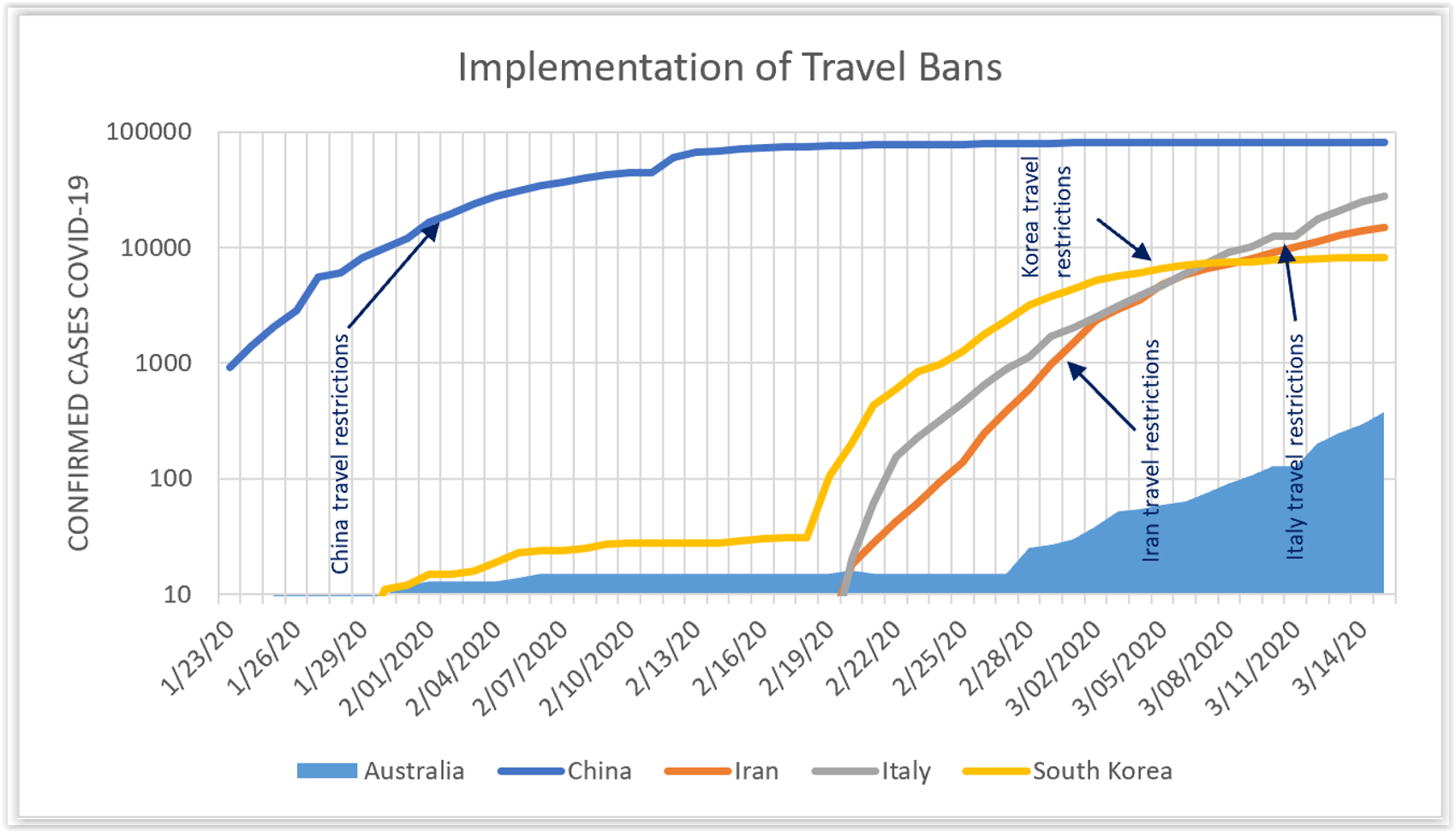}
\caption{Australia's implementation of travel bans on international travellers}
\label{fig:Figure 2.}
\end{figure}

Throughout this period, Australia’s response to the pandemic was coordinated at the national level with key decisions being to ban travellers from China, Iran, Korea and then Italy respectively. Fig. \ref{fig:Figure 2.} shows the timing of these travel bans in relation to the number of cases confirmed in Australia and these respective countries. The decision made by Australia to ban travellers from these countries was controversial, with the World Health Organisation (WHO) providing advice as late as the 29th February 2020 advising against the application of travel or trade restrictions to countries experiencing COVID-19 outbreaks\cite{RN105}. As a result of the advice provided by WHO, many countries delayed bringing in travel restrictions. But as of the 6th May, over 130 countries have bans on travellers from high risk regions or total border closures\cite{RN108}. 

Given the now widespread implementation of border controls, earlier implementation of international border controls for future pandemics could be considered. There is limited evidence regarding the effectiveness of international border controls with one review finding limited evidence for or against\cite{RN106}. Chinazzi et al looked at the effect of international travel limitations of COVID-19 through a disease transmission model\cite{RN107}. The number of case importations from China were found to be reduced by 80\% until the end of February, but the authors found that this would only delay the epidemic trajectory unless accompanied by public health measures to reduce transmission in communities\cite{RN107}. For island nations, another study found international travel restrictions to be a cost effective measure for mitigating a pandemic threat\cite{RN104}. However, more research is needed to evaluate the effect of ongoing international travel bans on the COVID-19 pandemic.

\subsection*{16th March - 20th April 2020: Time from State of Emergency Declaration in WA until first 0 new cases day}
Fig. \ref{fig:Figure 3.} demonstrates the growth in COVID-19 cases that occurs both locally in Western Australia and nationally in the whole of Australia. The period from the 15th March up until approximately the end of March shows high growth in the number of cases of COVID-19. The Australian response was rapid, with new social distancing measures implemented almost daily in response to the rising case load. The combination of measures locally and nationally has led to successful suppression of COVID-19 at the time of writing. 

Although it is evident that the combination of measures has led to COVID-19 suppression in Australia, it is difficult to identify the relative effects of different measures on the growth of cases in Australia.  There is a well-recognised time lag between the implementation of public health measures and a corresponding change in the growth of cases. The implementation date of key measures has been moved forward 7 days in Fig. \ref{fig:Figure 4.} below in order to better visualise the point on the curve at which these measures have started to take effect. 

Fig. \ref{fig:Figure 4.} below also plots Case Growth 5day average, which is a ratio between the new cases over the last 5 days to the new cases in the preceding 5 days. This is calculated with the following equation:

\begin{eqnarray} \label{eq1}
({\rm Case\quad Growth\quad 5-day\quad average}) & =  & \frac{n_t+n_{t-1}+n_{t-2}+n_{t-3}+n_{t-4}}{n_{t-5}+n_{t-6}+n_{t-7}+n_{t-8}+n_{t-9} } \\
\nonumber n & = & ({\rm new \quad cases \quad on \quad any \quad given \quad day}) \\
\nonumber t & = & ({\rm the \quad current \quad day})
\end{eqnarray}

Although this equation is not computing the effective reproductive number, $R_{eff}$, it does provide a different way of visualising growth in cases over time. A value more than 1 will see the number of new cases increasing, while a Case Growth value less than 1 will see new cases falls leading to suppression. A 5-day average was calculated to account for the daily swings in values. 

Fig. \ref{fig:Figure 4.} shows that the Case Growth 5-day Average continues to increase up to a maximum of 5 on the 24th March with measures such as mandatory self-isolation and bans on outdoor gatherings exceeding 500 people having an insufficient effect on the spread of the virus in Western Australia. The effects of a ban on indoor gatherings of more than 100 people and the borders being closed to foreigners can be seen from the 25th and 26th of March respectively. From the 25th March, the Case Growth 5-day average falls from near its peak of 5 down to a value of 1.11 on the 30th March. As of the 27th March, 83.1\% of Western Australia’s COVID-19 cases were acquired overseas with only 16.9\% the result of local transmission\cite{RN94}. It is possible that the drop in Case Growth reflects the effect of closing the borders, because overseas acquisition continues to be Western Australia’s largest source of infection.

Also, of note is that the Case Growth 5-day Average did not drop significantly below 1 until Australia went into Stage 3 restrictions. The effect of WA’s soft border closure, non-essential business closures and Australians being encouraged to work from home should be visible by the end of March with the Case Growth 5-day Average sitting at approximately 1 at this time. It was not until the implementation of gatherings being limited to two people that the Case Growth 5-day average fell significantly below 1 (0.54 on the 5th April). It is difficult to know how to interpret these findings without comparing to other states and countries. However, it is evident that there is a significant difference in the lockdown measures and hence the economic impacts of pursuing a mitigation strategy (reducing $R_{eff}$) as compared to a suppression strategy (keeping  $R_{eff}$ below 1). The question for the future is what level of normalcy can be achieved while still allowing for governments to keep control of any future outbreaks.

\subsection*{20th April - 15th May 2020: First 0 case day until first day with no active cases in Perth hospitals}
Western Australia recorded its first 0 new case day on the 20th April and since then has seen a sustained period of low numbers. As Fig. \ref{fig:Figure 5.} shows, from the 20th April until the 15th May, the number of active cases in the states has dropped from 103 to 7. The successful suppression of the epidemic in Western Australia has led to the development of a four-phase roadmap of easing social distancing restrictions\cite{RN109}. On the 27th April, Western Australia entered Phase 1, easing its first social distancing restrictions, allowing indoor and outdoor gatherings of up to 10 people. Further easing of restrictions occurred with Phase 2 beginning on the 18th May, that allows indoor and outdoor non-work gatherings up to 20 people, and cafes, restaurants and pubs to serve up to 20 patrons at any one time. The phases are designed to be implemented several weeks apart to allow time to measure the effect of easing these restrictions. 

There is significant uncertainty associated with what will happen to spread of the virus in Western Australia as different social distancing measures are eased. The presence of only 7 active cases in the state (as of the 15th May) combined with both interstate and international border closures will limit the number of new cases in the state. However, modelling of the epidemic in Perth using networks by Small and Cavanagh suggests that less than 80\% compliance with physical isolation measures risks further spread of infection\cite{RN110}. In a statement on the 16th April, The Australian Health Protection Principal Committee (AHPPC) stated that modelling is continuing to be carried out by University of Melbourne (Doherty Institute) pandemic modelling team to be able to, among other things, estimate what happens in the next fortnight\cite{RN111}. The accuracy and applicability of such models needs to be considered given the assumptions and uncertainties associated.

Regardless of modelling, both state and national governments appear to be aware of the increased risks associated with easing restrictions. On the 1st May 2020, the Prime Minister of Australia Scott Morrison outlined 15 precedent conditions that needed to be met for social distancing measures to be eased including testing and contact tracing capacity, adequate personal protective equipment, and surge capacity in the healthcare system\cite{RN112}. When the nation is as prepared as it can be for managing future outbreaks, restrictions can be eased with an attempt at a recovery.

Australia's growth slowed 0.3\% in the March quarter\cite{RN163}, with Australia expected to enter a recession when the June quarter statistics are released. This would be Australia's first recession in 29 years and is an indication of the effect COVID-19 is having on the Australian economy. However, the situation in Australia could be worse, especially when you compare Australia's performance to that of other countries. China's economy, where the virus originated, shrunk 6.8\% in the first quarter, the United States 5\% and across the European Union, 3.5\%\cite{RN164}. Australia's direct fiscal response to the pandemic has been the greatest among advanced economies, equating to 10.6\% of GDP, compared to 6.9\% in the United States and a median of 2.7\% across advanced economies\cite{RN165}. Although this has probably played a role in Australia's comparatively strong economic performance, Australia's successful suppression of pandemic has likely also played a role.

\section*{Further Analysis}

\subsection*{After Action Review }

As part of analysing the Western Australian response to the COVID-19 pandemic, an After Action Review (AAR) was carried out with colleagues from Integrated Energy Pty Ltd who had been involved in the COVID response, using the pro forma included in the appendices (Tables \ref{Table 1.}, \ref{Table 2.}, \ref{Table 3.} ). Several learning points came out of this exercise that can be implemented in the event of future outbreaks. 

The review recognised that Western Australia and Australia as a whole, had exemplary community adherence and buy-in to the public health measures that were implemented to ‘flatten the curve’ of the pandemic. Several contributing factors were identified as reasons for the community adherence to measures which resulted in significant restrictions in personal freedoms. These included consistent messaging from government at state and national levels, provision of economic support measures, and a clear plan to `flatten the curve’ that was supported by modelling showing the effect of community adherence. Australia was also lucky in that it was not the first Western country to experience a significant outbreak of the COVID-19 pandemic. Most of the social distancing restrictions in Australia were brought in the week of the 16th March, by which time Italy had already recorded more than 27,000 confirmed cases and over 2000 deaths\cite{RN113}. Witnessing the impact that COVID-19 had in another Western country may have had a positive affect on the Australian public's willingness to follow restrictions. 

The reports coming from Italy may also have led to the rapid response at both a national and state level and this was further supported by the Australian government activating the National Coordination Mechanism (NCM) on the 5th March 2020\cite{RN114}. This led to a centralisation of decision making and greater cooperation between national and state levels of government with the formation of a National Cabinet. The national cabinet comprises the prime minister and all territory premiers and chief ministers and is supported by the chief medical officers, who meet as the Australian Health Protection and Principles Committee (AHPPC). This mechanism has accelerated response time during the pandemic and allowed issues to be addressed as they emerged. 

The quarantine and isolation protocols that have been implemented in Western Australia have been successful in reducing case load in Western Australia. The setup of a government quarantine hub on Rottnest Island and the use of selected hotels for quarantine in Perth have been integral in removing infectious cases from the community. Pairing of these protocols with extensive, targeted testing has meant that Western Australia has identified almost all positive cases of the virus and removed them from the community early. The increase in contact tracing capability in Western Australia and nationally has led to faster identification of possible positive COVID-19 cases and removal of these cases from the community before continued community spread can occur. The combination of these measures is necessary to slow the spread of COVID-19. 

The national government has identified contact tracing as integral to slowing the rate of new infections and has worked to develop the contact tracing app COVIDSafe. COVIDSafe was available to download from the 26th April 2020 and works by recording 'digital handshakes' with other phone users within 1.5m\cite{RN160}. The app allows for faster and more complete contact tracing but its success relies on uptake by a significant number of community members. As of the 15th May 2020, there have been 5.7million downloads of the COVIDSafe app, equating to approximately 30\% of smartphone users in Australia over the age of 14\cite{RN161}. As both phones need the app for a handshake to occur, greater uptake is needed for the app to make a significant difference to contact tracing in Australia.

Australia and other island nations have a distinct geographical advantage that can be utilised and should be considered when implementing pandemic control measures. One study found that in New Zealand, border closures for a severe pandemic threat are a cost effective measure\cite{RN104}. Given Australia is also an island nation, this could be applicable in Australia, but more work needs to be done to evaluate Australia’s pandemic control measures in an economic sense. Australia has so far been successful in their pursuit of suppression of the pandemic, but there are many unknowns as to whether this is the most effective economic path. 

\section*{Social Context and Utility}

\subsection*{Epidemiological Modelling}

This epidemiological modelling (Small and Cavanagh \cite{RN110}) has its origins in the commencement of the COVID-19 outbreak in Perth, Western Australia\cite{RN110} and has continued to be developed.  The model was first used in a pandemic response workshop for a city of 100,000 people, where it served to demonstrate the dramatic range of outcomes which were possible, depending on the behaviour of constituents of the city, and degree of social distancing achieved.  This proved very effective in enabling appropriate action, both in the workshop and afterwards.  Subsequently the results were shared on professional social media, and an online conference, influencing thousands more.

The actual number of cases within Western Australia and nationally has also been plotted against the model with differences between predicted and actual discussed with state scientific authorities (Fig. \ref{fig:Figure 10.}). This enabled constructive discussion about the correlation between different public health measures and the outcomes that were achieved. 

The model has been further developed to explore what can be expected to happen as social restrictions are eased in Australia and the contact tracing app, COVIDSafe is rolled out (Fig. \ref{fig:Figure 6.} and \ref{fig:Figure 9.}. By modelling the effect of easing different measures on anticipated cases, all levels of government can make informed decisions on how to navigate an economic recovery while mitigating the risk of future outbreaks. Our modelling, for Australian consultancy company Integrated Energy, has specifically looked at the the effect of mass gatherings on the rate of spread of coronavirus with illustrations published in both the Wall Street Journal and The Australian\cite{RN162}. Network-based models first described in \cite{RN110} were applied to model the effects of various control measures on the rate of spread of infection. The results of this analysis, as depicted in Fig. \ref{fig:Figure 6.}, provide simulations under conditions including: no restriction, limits or bans on mas gatherings, effective contact tracing and shelter-in-place. 

While Fig. \ref{fig:Figure 6.} and \ref{fig:Figure 9.} provide a vivid portrayal of the relative merit of various generic control strategies, our modelling process can also be used to fit specific data driven scenarios. In Fig. \ref{fig-m30} we build models from the epidemiological history of Perth, up to March 30. From that date we apply prediction based on either extremely high adoption of social distancing ($80\%$ to $90\%$), or  lock-down. Under ``lock-down'' we assume no transmission in a proportion of the population, social distancing reduces transmission to diffusion within communities \cite{RN110}. Surprisingly, the observed data is not consistent with the social distancing scenarios --- even with very high levels of compliance. Conversely, model simulations of lock-down scenarios show very good agreement with the data (lower green bar and $90\%$ confidence interval). This simulation indicates that the work-from-home paradigm implemented in Western Australia had the effect of entirely eliminating transmission for a significant portion of the population.

\begin{figure}[!ht]
\centering
\includegraphics[width=0.9\textwidth]{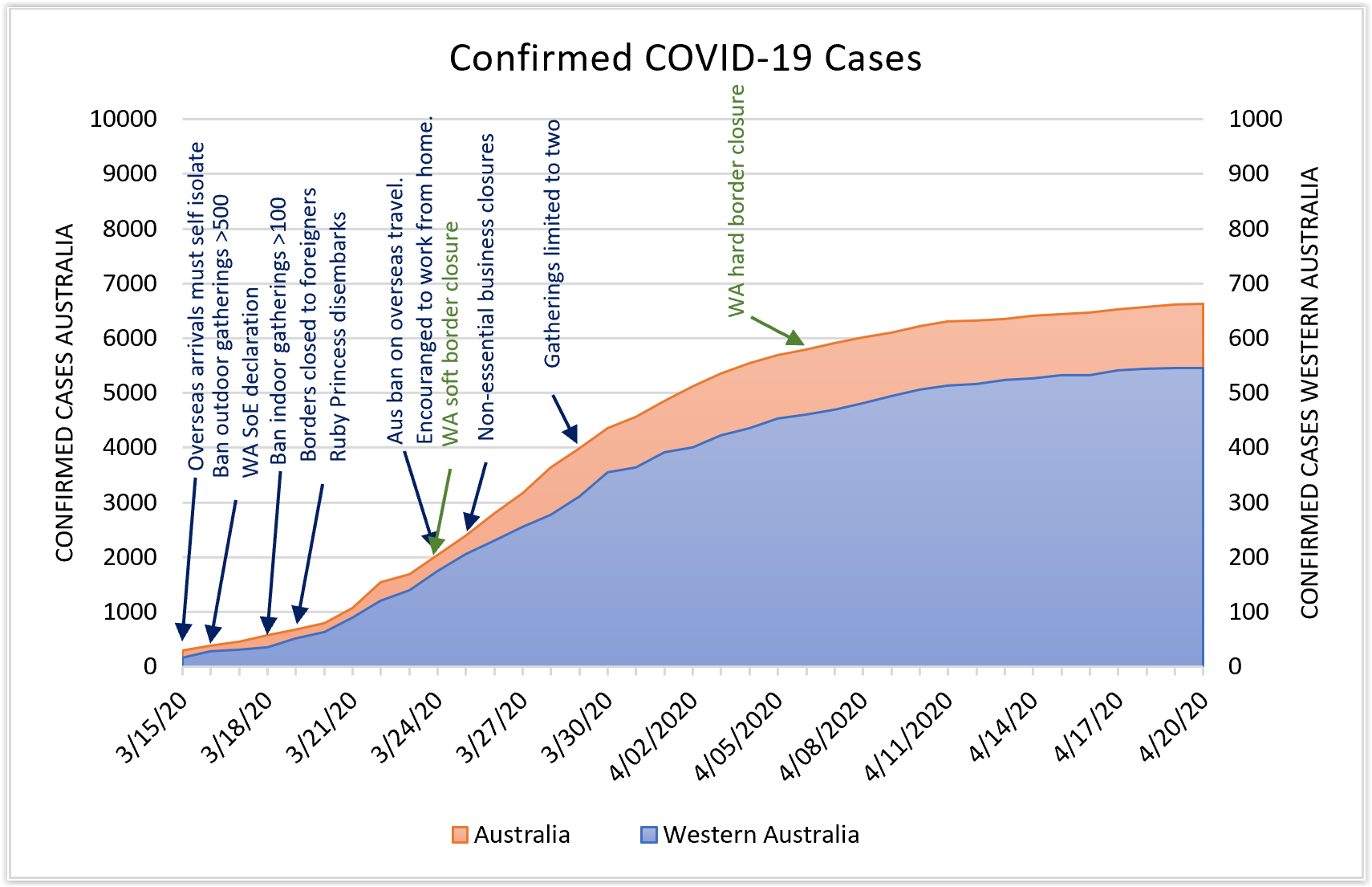}
\caption{Confirmed COVID-19 Cases in West Australia and Australia with the implementation of key public health measures from the 16th March 2020 until the 20th April 2020}
\label{fig:Figure 3.}
\end{figure}

\begin{figure}[!ht]
\centering
\includegraphics[width=0.9\textwidth]{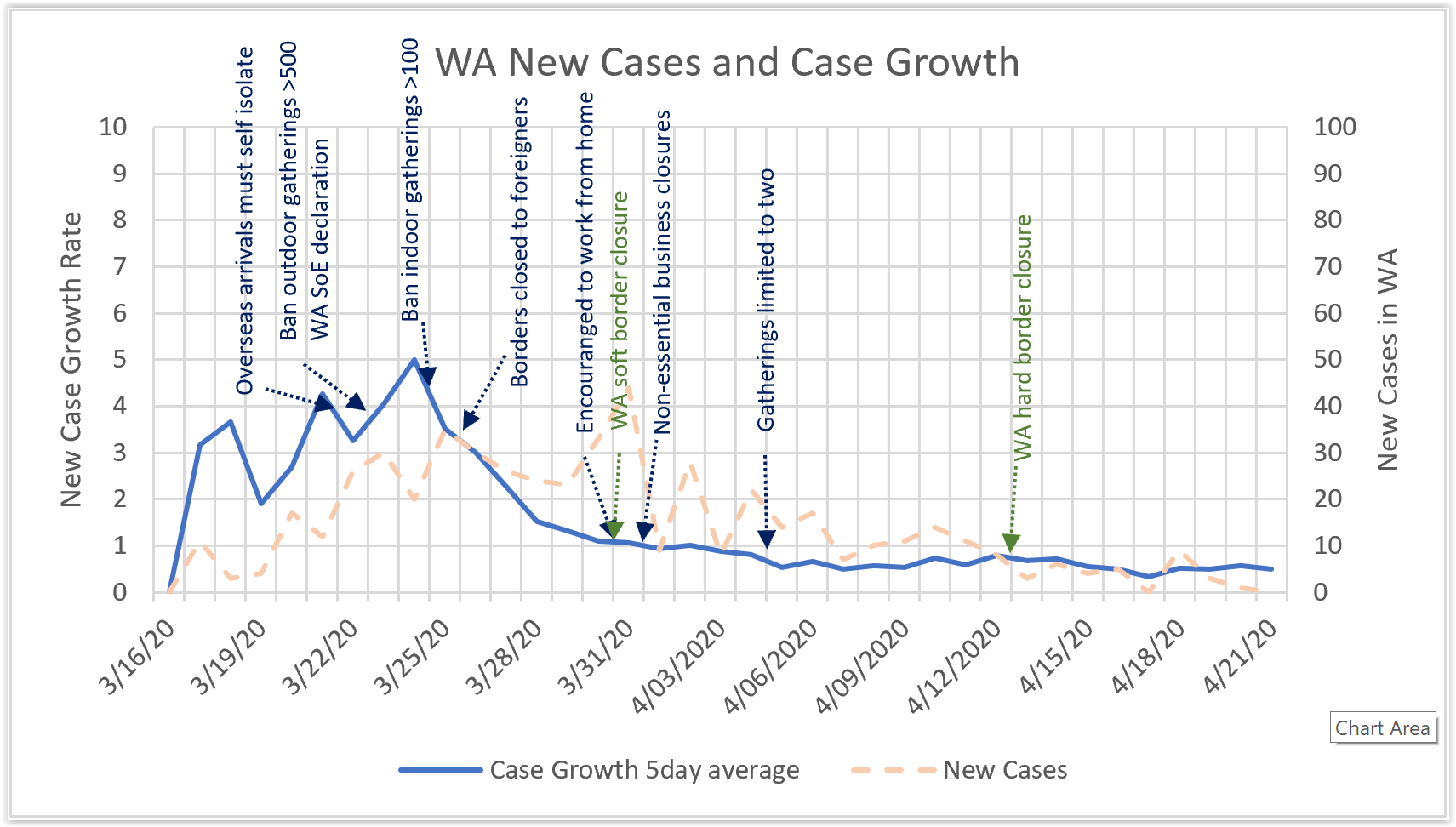}
\caption{West Australia Calculated Case Growth and New Cases from 16th March 2020 until the 21st April 2020}
\label{fig:Figure 4.}
\end{figure}

\begin{figure}[!ht]
\centering
\includegraphics[width=0.9\textwidth]{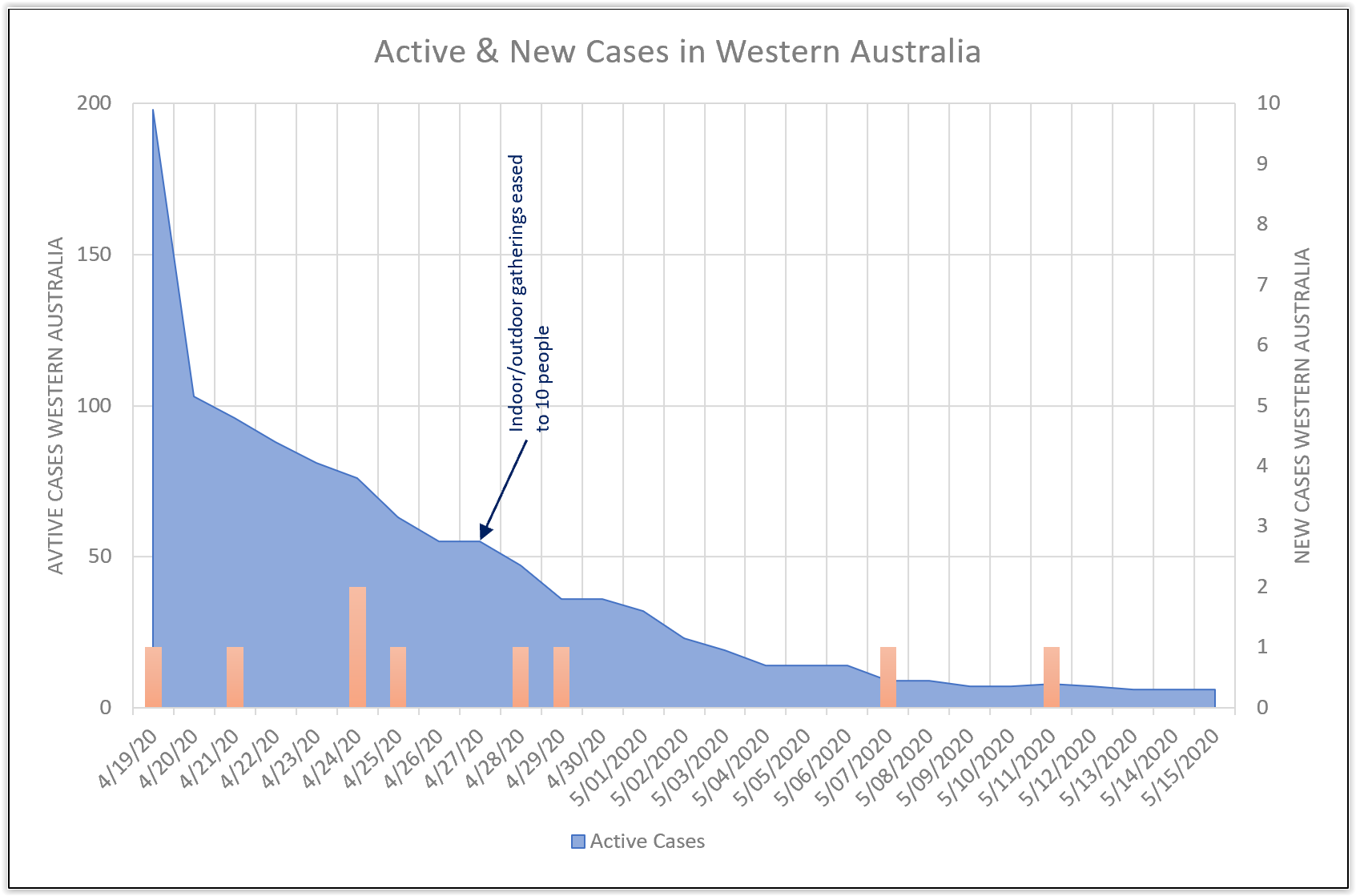}
\caption{Active and New Cases in West Australia between the 19th April 2020 and the 6th May 2020}
\label{fig:Figure 5.}
\end{figure}

\begin{figure}[t]
\centering
\begin{tabular}{c}
(a) \\\includegraphics[width=0.8\textwidth]{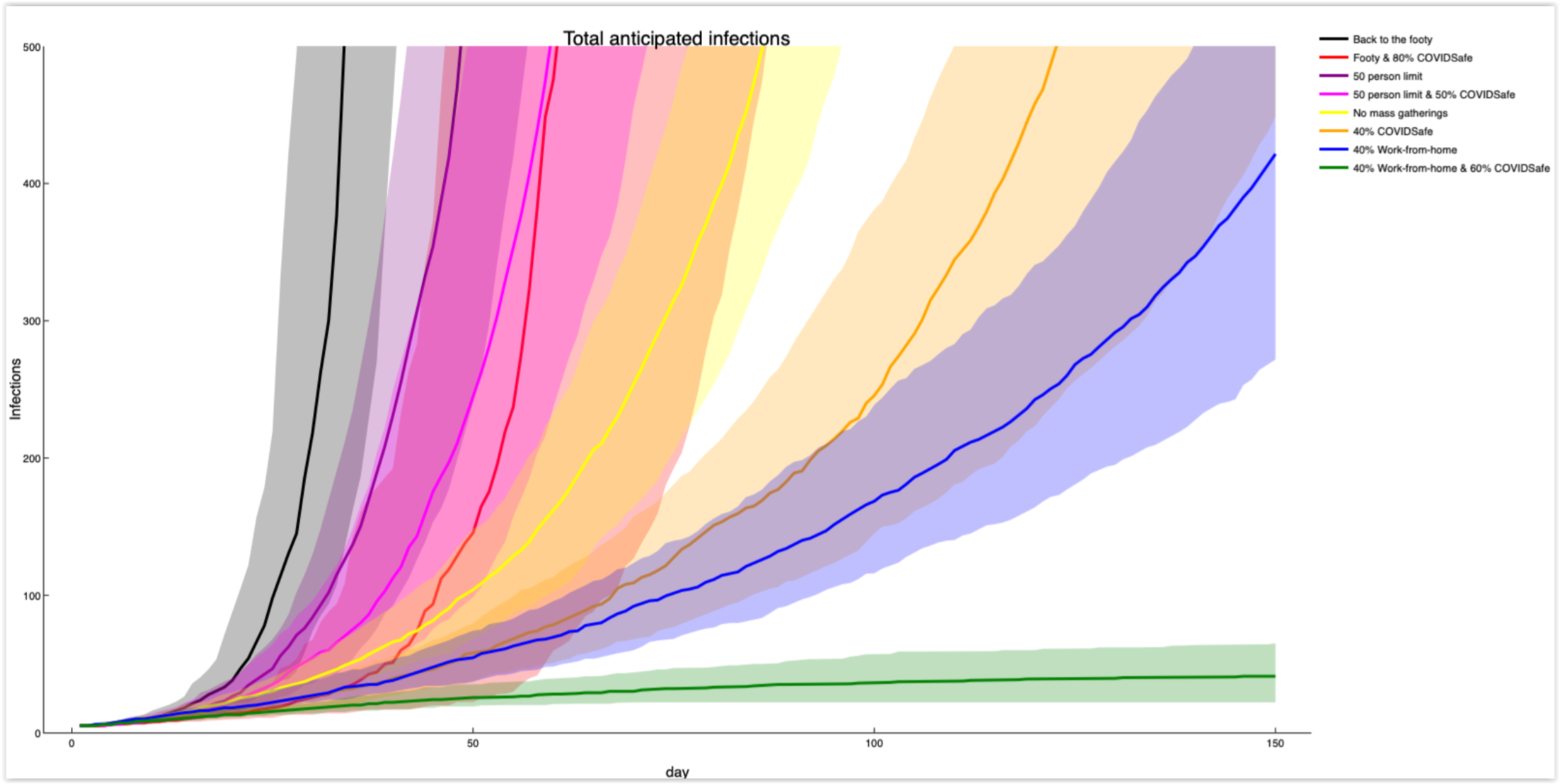}\\
(b) \\ \includegraphics[width=0.8\textwidth]{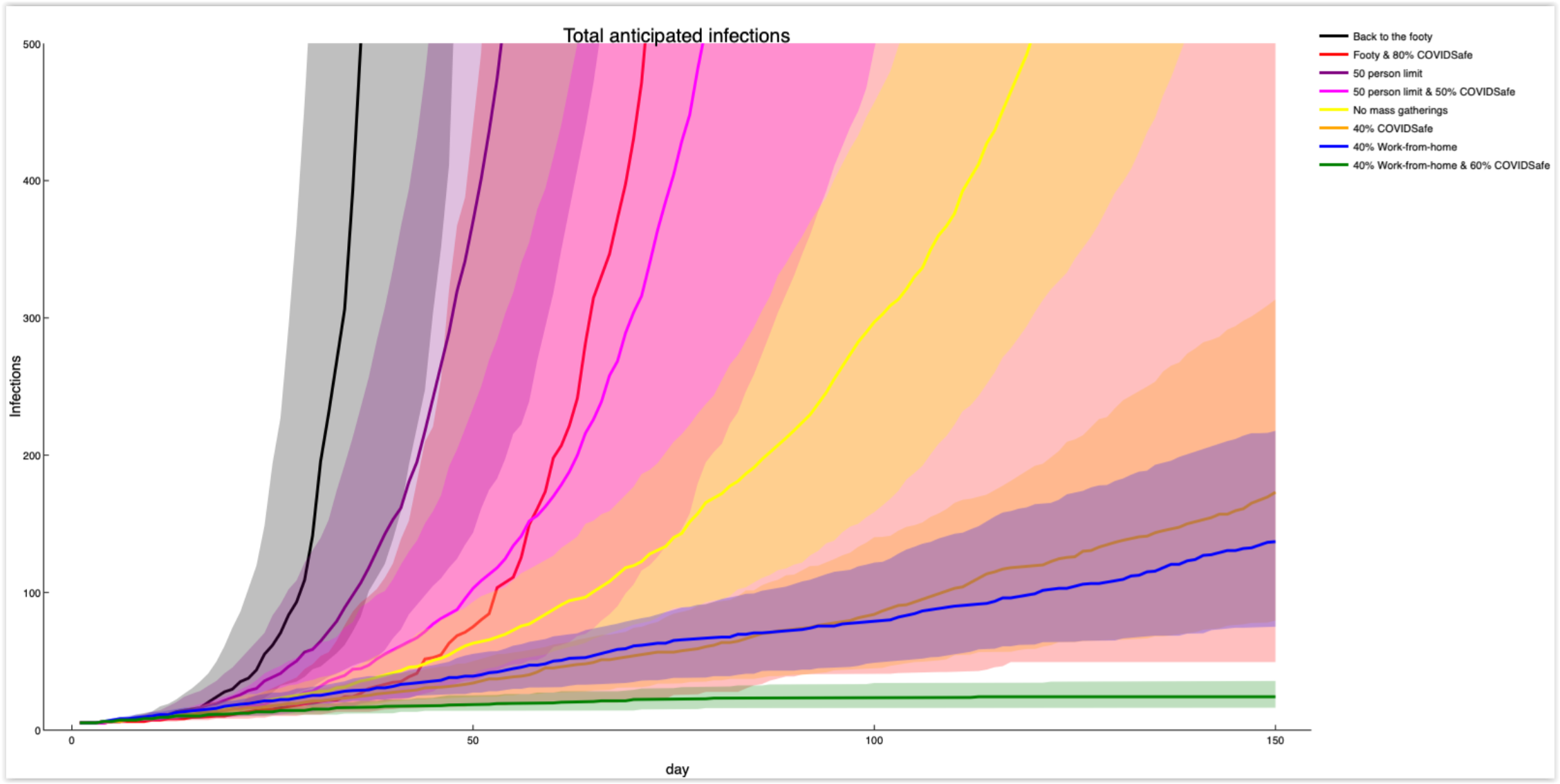}\\
(c) \\  \includegraphics[width=0.8\textwidth]{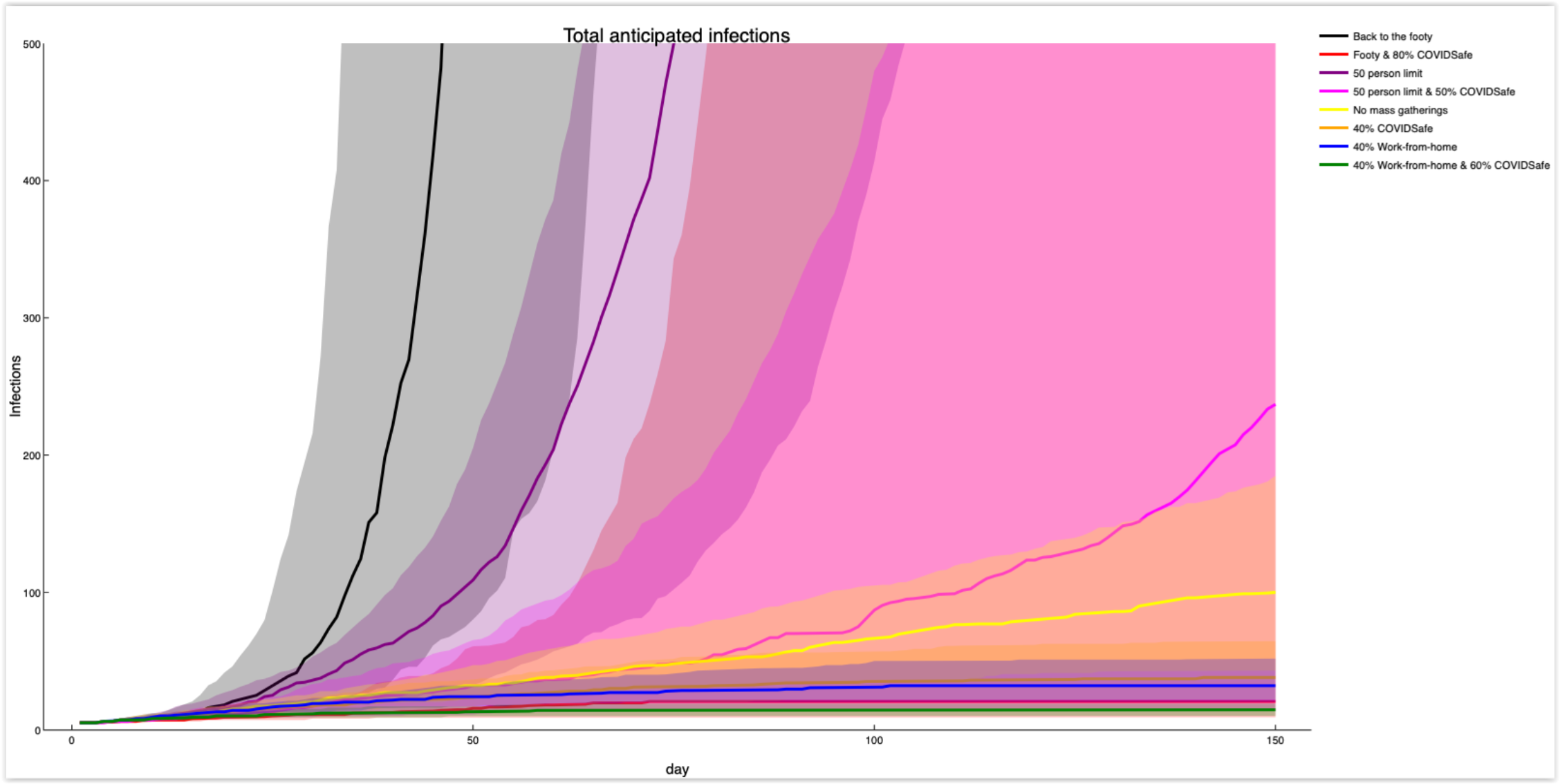}
\end{tabular}
\caption{Total anticipated infections for different public health restrictions assuming (a) p = 1/9 and r = 1/7; (b) p = 1/10 and r = 1/6; and (c)  p = 1/10 and r = 1/4}
\label{fig:Figure 6.}
\end{figure}

\begin{figure}[ht]
\centering
\includegraphics[width=0.9\textwidth]{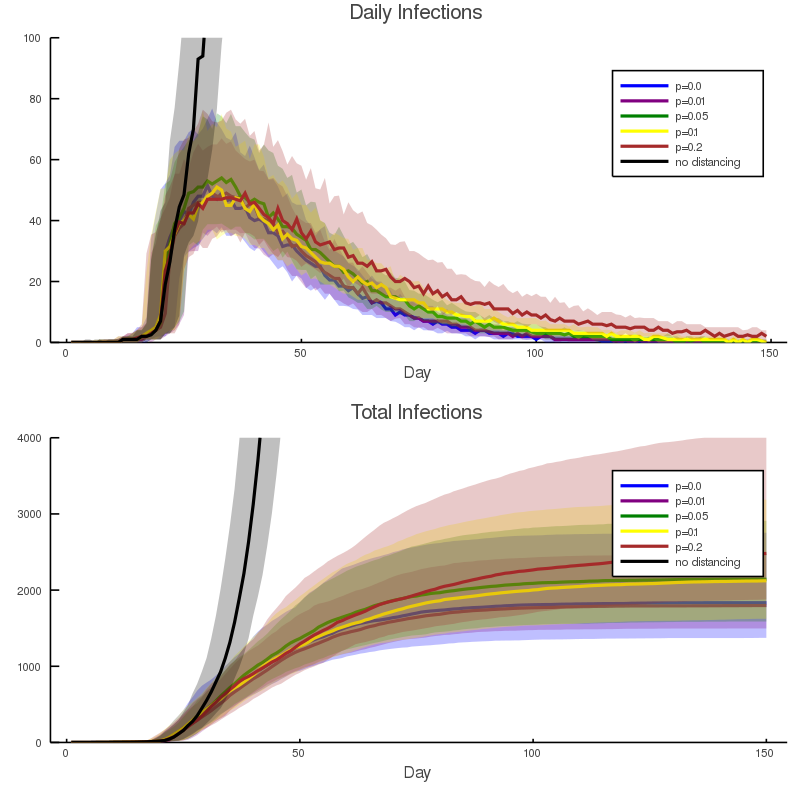}
\caption{Predicted daily and total infections for different adherence to physical distancing measures}
\label{fig:Figure 9.}
\end{figure}

\begin{figure}[ht]
\centering
\includegraphics[width=0.9\textwidth]{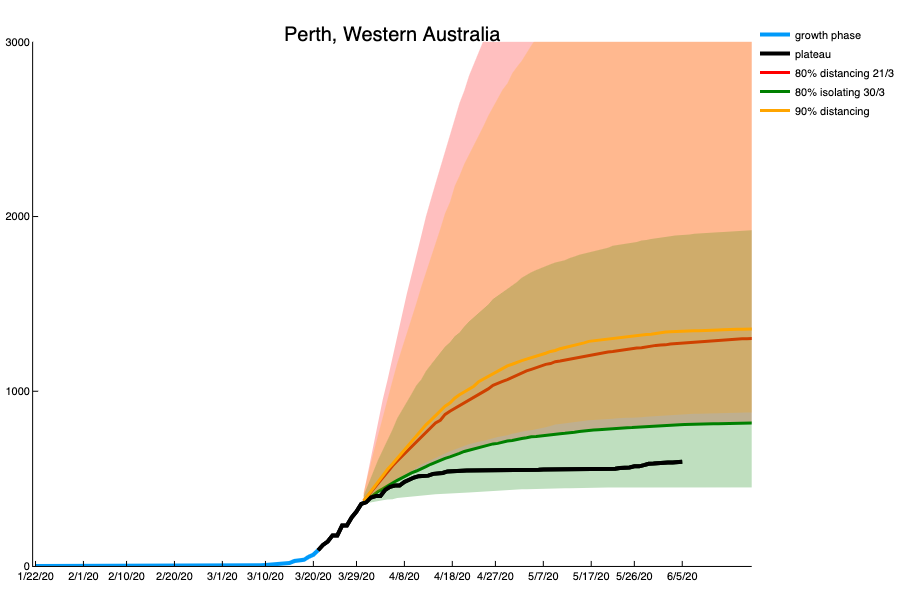}
\caption{Effect of ``lockdown'' and social distancing practises for Perth}
\label{fig-m30}
\end{figure}

\begin{figure}[h!]
\centering
\begin{tabular}{cc}
\includegraphics[width=0.45\textwidth]{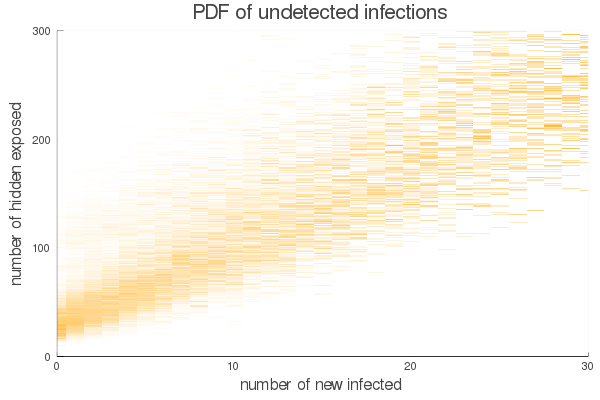} &
\includegraphics[width=0.45\textwidth]{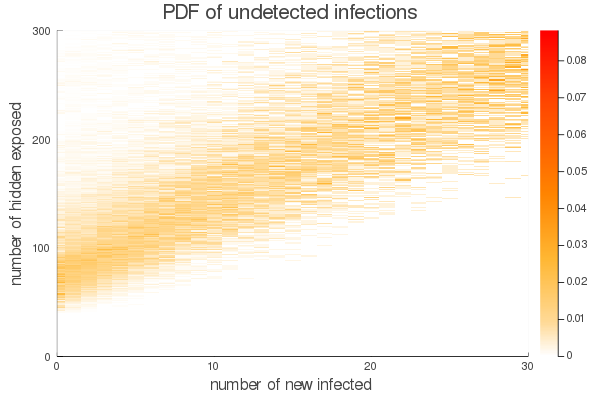}\\
(a) & (b)
\end{tabular}
\caption{Estimate of undiagnosed cases for (a) Perth and (b) Melbourne. The horizontal axis is the number of detected new diagnoses of COVID-19 on a given day, the vertical axis is the estimated distribution of the number of remaining exposed (but asymptomatic and undiagnosed) cases still in the community. Note the approximately linear trend, wide variance and significant level of remaining community infection even once new diagnoses are reduced to $0$. An expected value of about $25$ for Perth and $80$ for Melbourne.}
\label{fig9b}
\end{figure}

\begin{figure}[t]
\centering
\includegraphics[width=0.9\textwidth]{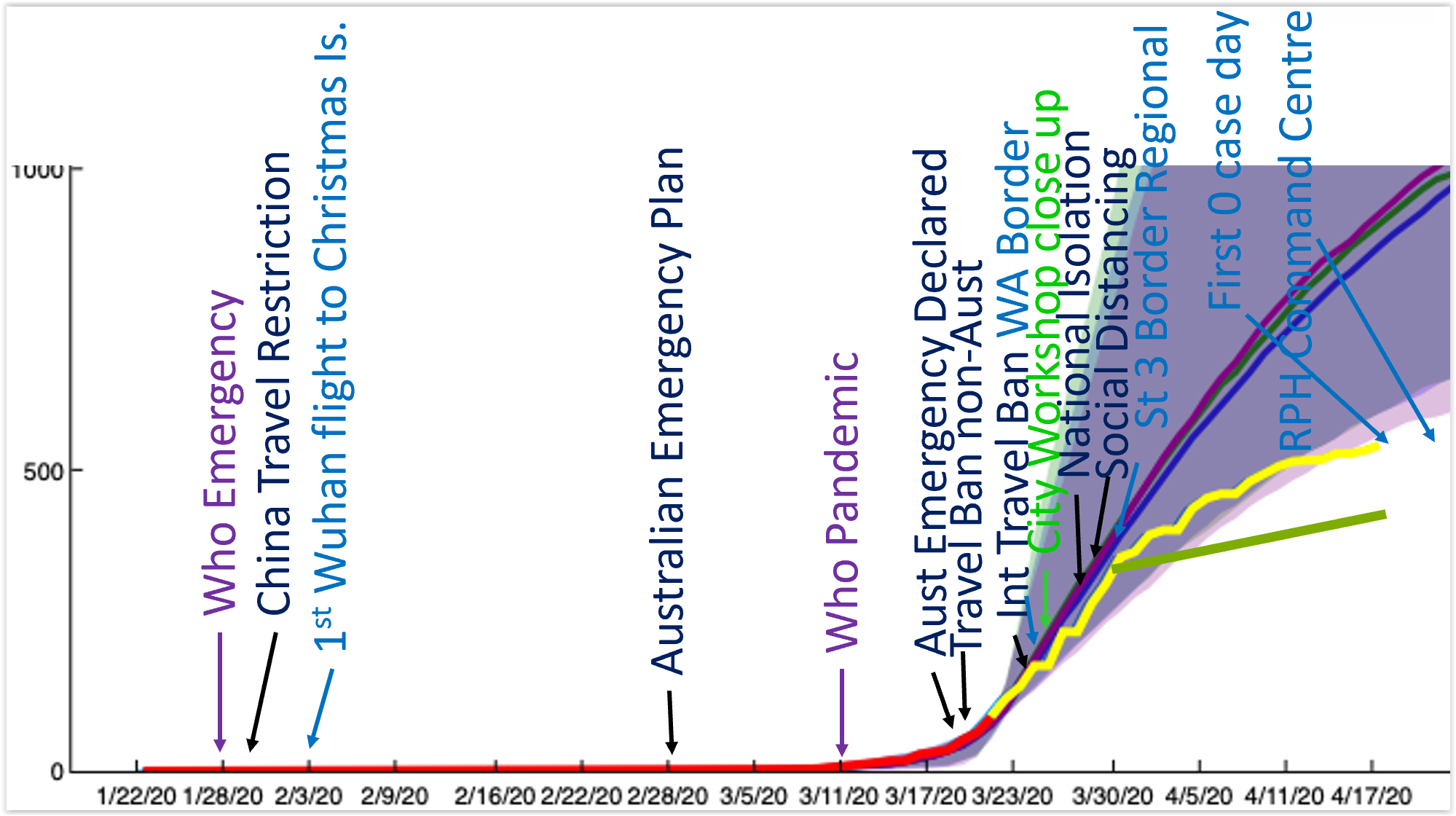}
\caption{ Actual and Modelled WA Cumulative COVID-19 Cases and the implementation of public health measures}
\label{fig:Figure 10.}
\end{figure}

Figure \ref{fig9b} illustrates the risk posed by undiagnosed cases. These computations provide an estimate of the remaining community infections in the tail of the Australian outbreaks for a given number of new infections. Using the modelling structure described in \cite{RN110} we estimated biologically plausible epidemic parameters and computed epidemic outbreak trajectories which matched (via maximum likelihood) the observed time series data up to May 20. The model structure was constructed to capture the various macroscopic controls imposed on the community as illustrated in Fig. \ref{fig:Figure 3.}. We then continued these simulations forward in time, taking note of the number of new infections and the corresponding number of current exposed individuals in the communities. We repeated this for $500$ simulations over $180$ days. Days on which $0$ new infections were observed on days immediately subsequent to other days with $0$ recorded infections were ignored (that is, the resulting distribution shown here is only for the first ``$0$-day'' in any string of $0$-days). From this we computed histograms reflecting the probability distribution of the number of extant exposed individuals given an arbitrary number of newly recorded infections. 

Our results show an approximate linear trend with wide variance. More importantly, when $0$ new infections are recorded we estimate approximately (median) of $25$ community infections (but asymptomatic) for Perth, and about $80$ for Melbourne. This reflects a strong risk of a second outbreak even once zero detected infection is achieved. This resurgence from undetected community infection is precisely the scenario currently facing Melbourne.

As of the 20th May 2020, Western Australia has almost eradicated COVID-19, however temporarily, with only 3 active cases and 1 patient hospitalised across the state\cite{RN115}.

\subsection*{COVID-19 Business Safety Course and Health Care Evolution}
An effective and fundamental tenet of the Australian and West Australian COVID-19 response has been to cascade the required accountability for changing behaviour down to individual businesses and business premises, with a requirement for any businesses which had been required to close, to develop a COVID Safety Plan, and then reopen with their operations in accordance with the plan.  To do this most effectively required some knowledge of infectious disease management, risk management, and the specifics of COVID-19.  Consequently a live short course was developed to enhance this knowledge amongst small businesses, with supporting consulting services appreciated particularly where clients had an international spread of operations.

Health care systems were forced to re-engineer in real time for a threat for which they were not designed,  Hospital based COVID clinics were introduced in the second week of March 2020 in Western Australia, enabling cohorting of walking (vertical) patients in dedicated areas within hospitals to facilitate safe assessment and rapid access to diagnostic testing.  Emergency Departments largely reconfigured into "hot"(horizontal potential COVID patients) and "cold"(COVID unlikely) areas to keep potentially or actually infective patients away from those unlikely to be infected.  Common clinical practices - non invasive ventilation, for instance - became limited due to their potential to generate aerosols.  Penetration of COVID positive patients into general ward environments and intensive care units remained at relatively modest levels.  Activity in Emergency Departments was generally very low, leading to lower levels of ambulance ramping - an unanticipated benefit of low levels of COVID and high levels of concern in the broader community.

Ambulance also noted generally low levels of activity, high levels of concern amongst frontline staff, and tasking and clinical guideline modifications to protect frontline staff from disease transmission in the event of an interaction with a COVID positive patient, either known or unknown.

Whole of population telehealth services were introduced by the Australian Federal Government on 30 March 2020, enabling tele consultations for any Medicare eligible patient provided by a practitioner qualified to deliver that service under usual circumstances, provided it was feasible and safe to do so.  These changes included pivoting from face to face consultations to Medicare subsidised telehealth consultations.

\begin{figure}[t]
\centering
\includegraphics[width=0.9\textwidth]{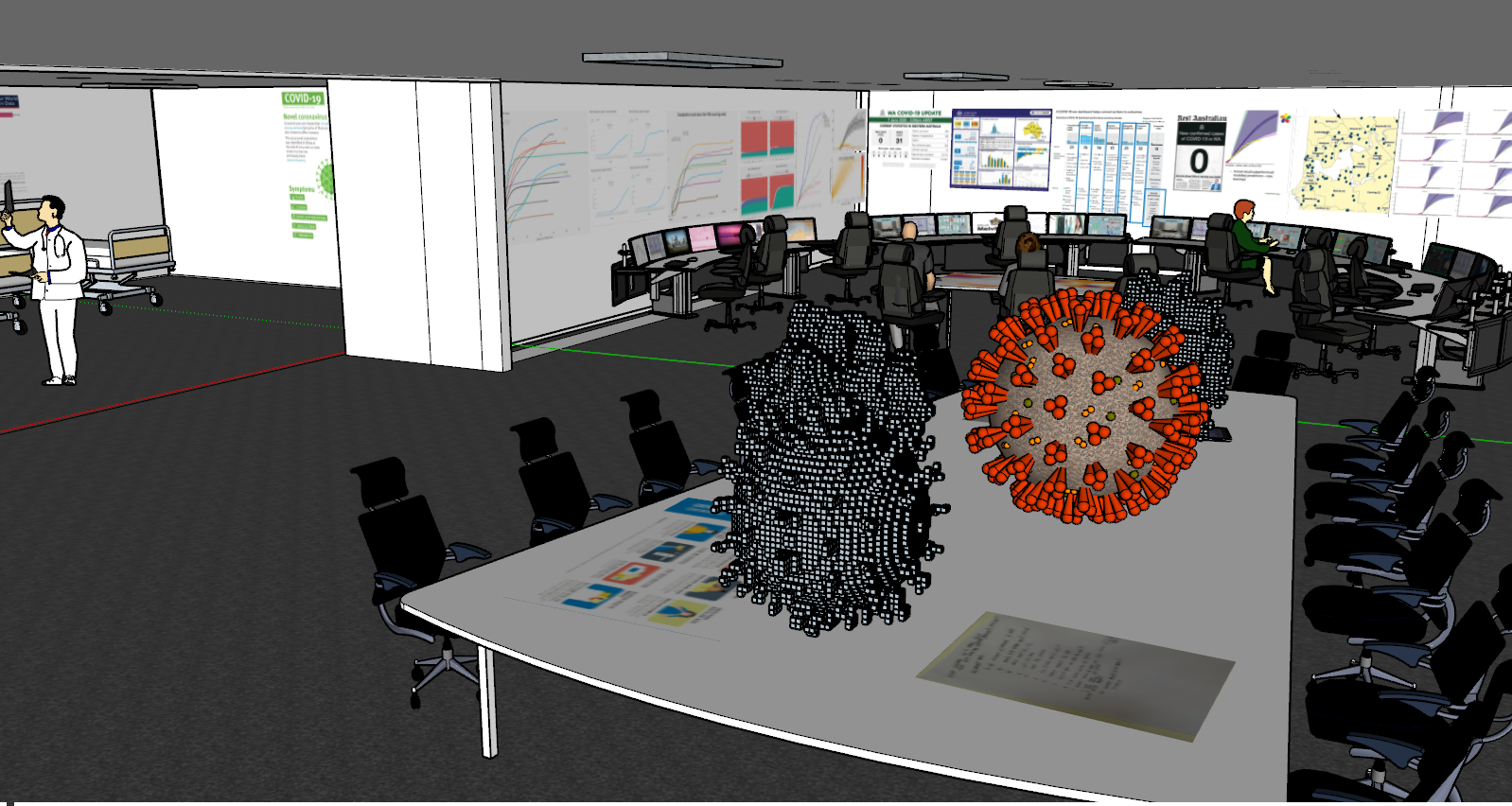}
\caption{ Representative Actual and Digitally Created Coronavirus Models for Training AI Algorithms}
\label{fig:Figure 10b.}
\end{figure}

As the pandemic progressed, the role of digital solutions, and integrated operations which could continue effectively over distance became apparent.  These models of working have been adopted widely in Western Australia over the last decade, particularly in the resource and energy industries\cite{RN166}, and allowed production to continue relatively uninterrupted through the pandemic, with Australia's trade surplus reaching an all time high.  Similarly, increasing application of AI techniques have been used, for example in Figure \ref{fig:Figure 10b.}.  This Figure shows a 3d model of the COVID-19 virus, alongside several other mathematically generated shapes which have some similar features, produced by Dr Andrew Marsh.  This data set is used to help train algorithms to automatically detect COVID-19 from a series of digital diagnostic images.

\subsection*{Best Practice Framework and Gap Analysis}
\begin{figure}[t]
\centering
\includegraphics[width=0.9\textwidth]{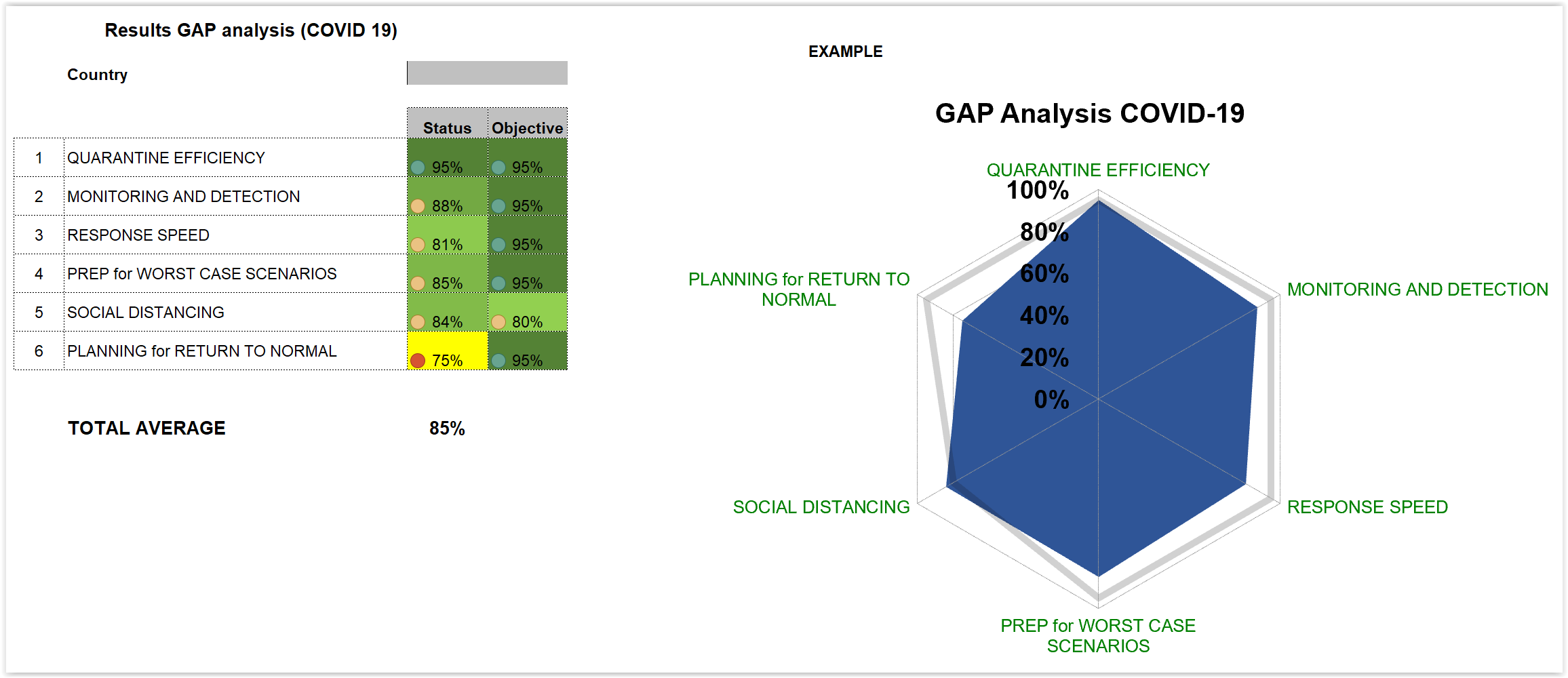}
\caption{Example of GAP Analysis tool used for visual communication of COVID-19 pandemic response}
\label{fig:Figure 11.}
\end{figure}

Through a combination of tracking the pandemic throughout this period, modelling of the epidemic with networks, and consultation, a Best Practice Framework and Gap Analysis has been developed. These tools allow for greater visual communication of the range of government responses that have been implemented globally. An example of the Gap Analysis is shown in Figure \ref{fig:Figure 11.}. 

\section*{Conclusions}

Regardless of the context or setting, it is important that steps are taken to learn from previous experience. Given the unprecedented health and economic impacts of COVID-19, this is of upmost importance when it comes to learning from the varying pandemic responses that have been implemented in different jurisdictions. Western Australia at a state level, and Australia at a national level, have so far demonstrated successful suppression of COVID-19. This success has been analysed in an attempt to distill information that can be applied to better respond to future outbreaks. 

The response to the pandemic in Australia was fast, coordinated effectively between different levels of government, had widespread community support, and was backed up with effective quarantine and isolation procedures, border closures and prompt expansion of testing capabilities. Economic impact has been limited by the implementation of one of the world's largest stimulus packages per capita, while innovation and working from home arrangement have allowed many industries to continue close to business as usual. Australia's largely successful suppression of the pandemic can also thank some country specific factors such as our relative isolation as an island nation.  Consequently Western Australia's experience could be considered the black swan of this black swan global event.

Australians watched the COVID-19 crisis evolve in other countries prior to the pandemic developing in Australia. Reports of the unprecedented impact of the pandemic in other countries, combined with Australia's recent experience with bushfires likely contributed to Australia's fast response and the exceptional community adherence shown by the public. This is the first pandemic to have occurred in the new age of social media and globalism.  Consistency in modelling, data analysis, and communication is paramount to assist our community response.

Australia's response to the pandemic has not been without its missteps either. Shortages of personal protective equipment led to some issues in primary and secondary care, and problems with cruise-ship arrivals led to a significant number of cases across Australia. These issues emphasised the importance of developing local supply chains for essential equipment and highlighted the importance of integrated communication and improved risk assessment for the future. Further modelling and communication as the pandemic continues to progress globally will be very important, as Australia responds to the continuing population health and economic challenges that remain.

\clearpage

\bibliography{sample}

\section*{Acknowledgements }

Data utilised for creation of Figures was obtained from \url{https://github.com/CSSEGISandData/COVID-19}.
This work was supported by Integrated Energy Pty Ltd.

\section {Author contributions statement}

All authors contributed to and reviewed the manuscript.

\newpage
\begin{table}[t]
  \caption{Completed After Action Review of WA COVID-19 Response. Participants; David Cavanagh, Paul Bailey, Cameron Ferstat, Peter Deveugle, Mark Hoey}
    \label{Table 1.}
    \begin{tabular}{|p{25.465em}|l|}
    \toprule
    \textbf{Plan} & \multicolumn{1}{p{25.465em}|}{\textbf{Actual}} \\
    Documented in WA COVID Pandemic Response Plan & \multicolumn{1}{p{25.465em}|}{Melbourne Uni started Covid modelling from 17 Jan} \\
    •        Pre-existing plan for flu pandemic updated for Covid-19 & \multicolumn{1}{p{25.465em}|}{Response Agency to review/revise plans from 2 Feb} \\
    •        First version 2006, SARS update 2014, Covid March 2020 & \multicolumn{1}{p{25.465em}|}{First patients suspected, not tested (travel criteria) 4 Feb} \\
    Prevention & \multicolumn{1}{p{25.465em}|}{Prevention \& preparedness were already past} \\
    •        Surveillance programs & \multicolumn{1}{p{25.465em}|}{Alert – 2 Feb, Aus travel restrictions - 1 Feb (China), 1 Mar (Iran), 5 Mar (Korea), 11 Mar (Italy)} \\
    •        Communicable disease prevention activities & \multicolumn{1}{p{25.465em}|}{Standby - 7 Feb} \\
    •        Establishment of infection control guidelines & \multicolumn{1}{p{25.465em}|}{Response} \\
    •        Screening at borders & \multicolumn{1}{p{25.465em}|}{•        Health Incident controller appt 14 Feb, PHEOC } \\
    •        Collaboration with national/international health agencies & \multicolumn{1}{p{25.465em}|}{•        Australian Grand Prix Cancelled March 12} \\
    Preparedness & \multicolumn{1}{p{25.465em}|}{•        National Cabinet Initiated 5 March} \\
    •        Establish plans/legislation to support an effective response. & \multicolumn{1}{p{25.465em}|}{•        WA State of Emergency – 16 Mar } \\
    •        Identify vulnerable populations and provide support. & \multicolumn{1}{p{25.465em}|}{•        Social distancing / business / travel constraints} \\
    •        Establish pharmaceutical and PPE stockpiles & \multicolumn{1}{p{25.465em}|}{•        Community adherence >98\%} \\
    Response & \multicolumn{1}{p{25.465em}|}{•        Australian Emergency – 20 Mar} \\
    •        Alert – monitor situation & \multicolumn{1}{p{25.465em}|}{•        Emergency Ops Centre – 24 Mar} \\
    •        Standby – consult with Communicable Diseases Network Australia and the AHPPC. & \multicolumn{1}{p{25.465em}|}{•        City of Melville Response Workshop 25 March, debrief on 9 April} \\
    •        Response – appoint incident controller, implement tactics  & \multicolumn{1}{p{25.465em}|}{•        PPE arrived, extra ICU capacity, extra testing} \\
    •        Flatten the Curve (March) & \multicolumn{1}{p{25.465em}|}{Attempt at Recovery:} \\
    •        Suppression / Possibility to Eradicate (April) & \multicolumn{1}{p{25.465em}|}{•        First easing of WA restrictions – 27 Apr} \\
    •        Staged Easing of controls (27 April WA)  & \multicolumn{1}{p{25.465em}|}{•        National roadmap to a COVID Safe Australia -3 steps – Reopening – Most Open – All at work } \\
    •        Provide Vaccination when available & \multicolumn{1}{p{25.465em}|}{•         WA 4 phase roadmap – COVID-19 Safety Plan a requirement } \\
    Attempt at Recovery (WA Roadmap) &  \\
    \bottomrule
    \end{tabular}%
\end{table}%

\begin{table}[htbp]
  \centering
  \caption{Completed After After Review of WA COVID-19 Response Continued}
\label{Table 2.}
    \begin{tabular}{|p{25.465em}|l|}
    \toprule
    \textbf{Differences Positive \& Why} & \multicolumn{1}{p{25.465em}|}{\textbf{Learning for Future}} \\
    Australia Acted before a global pandemic was called by WHO: & \multicolumn{1}{l|}{\multirow{4}[0]{7.5cm}{•Australia has distinct geographical advantages due to being an island. Therefore, worldwide WHO advice in regard to travel controls may not be the best course of action when the specifics of a country are considered. }} \\
    •        Australia acted on own medical experts’ advice. &  \\
    •        Australian geographic advantages (island continent) &  \\
    Exceptional community adherence to public health measures: &  \\
    •        Italy experience strongly influenced rapid and thorough adoption by Australian community & \multicolumn{1}{l|}{\multirow{3}[0]{7.5cm}{•High public goodwill has a narrow window and is dependent on government behaviour – government should act to encourages adherence (consistent messaging, economic support, clear plan, education) and optimise this time.}} \\
    •        Australian government (national/state/local) messaging was consistent. Also had social reinforcement &  \\
    •        Direct communication by Premier with social media &  \\
    •        Australians conditioned by bushfire experience & \multicolumn{1}{l|}{\multirow{3}[0]{7.5cm}{•Case load was reduced quickly through a combination of community adherence to restrictions, effective isolation of cases and early shutdown.}} \\
    •        Modelling used to show the impact of peoples’ choices &  \\
    •        Economic support &  \\
    Reduced case load more quickly than predicted: &  \\
    •        Exceptional community adherence to public health measures. & \multicolumn{1}{l|}{\multirow{4}[0]{7.5cm}{•The speed of the response in WA was critical to success. Recent bushfires and precedents set in countries such as Italy were factors but good governance at both state and national levels is also critical. Setup of National Cabinet body was key.}} \\
    •        Highly effective isolation of external cases &  \\
    •        Remote working/education widely adopted.  &  \\
    •        Extensive early shutdown. &  \\
    Speed of response in introducing social distancing/travel bans  &  \\
    •        Conditioned by bushfire experience & \multicolumn{1}{l|}{\multirow{4}[0]{7.5cm}{•Effectiveness of quarantine and isolation procedures in WA were integral to reducing case load. Setup of government quarantine hubs such as Rottnest Island and hotels in Perth have played a role in removing infectious cases from the community.}} \\
    •        Central decision making body (National Cabinet) &  \\
    •        Reports of pandemic from Italy, China, etc &  \\
    Quarantine/Isolation efficiency:  &  \\
    •        Use of Rottnest Island for effective quarantining. &  \\
    •        Emergency Powers and natural advantages & \multicolumn{1}{l|}{\multirow{3}[0]{7.5cm}{•The knowledge gained from the suppression of the pandemic in Western Australia is applicable to other jurisdictions if adapted to their individual contexts. }} \\
    •        Testing allocated objectively, then increased ASAP. &  \\
    •        Centralised decision making for pandemic response &  \\
    City level engagement rapidly effective &  \\
    •        Workshop delivered remotely which addressed spread of scenarios and risk, supported by modelling. & \multicolumn{1}{l|}{\multirow{2}[0]{7.5cm}{•A low mortality rate has been achieved in Australia through good healthcare facilities, standardised protocols and a large testing capacity.}} \\
    Low mortality &  \\
    •        Good standards of health facilities. Not overwhelmed. &  \\
    •        Good access (public health care) & \multicolumn{1}{l|}{\multirow{2}[0]{7.5cm}{•In Western Australia, cruise ships such as Artania were managed well and make up a limited number of cases in Western Australia. }} \\
    •        Standardised and current protocols  &  \\
    •        Less elderly demographic than some countries &  \\
    •        Large testing capacity to record asymptomatic cases & \multicolumn{1}{l|}{\multirow{3}[0]{7.5cm}{•High level of economic output has been maintained in some sectors through uptake of working from home and remote ops capability.}} \\
    Cruise ships were safely managed in WA &  \\
    •        Learning from Ruby Princess &  \\
    •        Commonwealth/State solution &  \\
    Economic output maintained in sectors such as mining and energy & \multicolumn{1}{l|}{\multirow{4}[0]{7.5cm}{•Necessity has been an efficient driver of change with an accelerated rate on innovation seen across industries in Western Australia (Telehealth, Royal Perth Operations Centre, working from home, Digital Learning).}} \\
    •        Declared an essential service &  \\
    •        Integrated/remote ops philosophy and capability &  \\
    Accelerated rate of innovation (eg Telehealth, Royal Perth Ops Centre, Working from Home, Digital Learning): &  \\
    •        Necessity driver of change and people open to it. & \multicolumn{1}{l|}{\multirow{4}[0]{7.5cm}{•Australia has led the way in terms of the amount of economic stimulus provided to keep the economy afloat. Rapid suppression of the virus will likely lead to a more positive economic outcome, with faster easing of public health measures.}} \\
    Reduced environmental impact: &  \\
    •        Less commuting, industrial output  &  \\
    Economic stimulus and likely overall economic outcome: &  \\
    •        Rapid suppression enables faster easing of restrictions. &  \\
    WA as a centre of innovation/research: &  \\
    •        Skilled, stable workforce, developed information systems & \multicolumn{1}{l|}{\multirow{3}[0]{7.5cm}{•WA’s success suppressing the virus, combined with its skilled workforce and IT systems, means knowledge can be shared effectively to other places.}} \\
    WA as a centre of knowledge that can share services: &  \\
    •         Good result, structured learning, skilled resources. &  \\
    \bottomrule
    \end{tabular}%
\end{table}%

\clearpage

\begin{table}[t!]
  \centering
  \caption{Completed After After Review of WA COVID-19 Response Continued}
 \label{Table 3.}
   \begin{tabular}{|r|l|}
    \toprule
    \multicolumn{1}{|p{25.465em}|}{\textbf{Differences Negative \& Why}} & \multicolumn{1}{p{25.465em}|}{\textbf{Learning for Future}} \\
    \multicolumn{1}{|p{25.465em}|}{PPE Shortages:} & \multicolumn{1}{l|}{\multirow{3}[0]{7.5cm}{•Shortages of PPE led to less than ideal practices at hospitals in WA. Ensuring larger stockpiles and developing local supply chains would mitigate future risk.}} \\
    \multicolumn{1}{|p{25.465em}|}{•        Supply chains dependent on international supplies} &  \\
    \multicolumn{1}{|p{25.465em}|}{•        Insufficient local stockpiles for pandemic} &  \\
          &  \\
    \multicolumn{1}{|p{25.465em}|}{Handling of Ruby Princess cruise-ship (occurred in New South Wales):} & \multicolumn{1}{l|}{\multirow{3}[0]{7.5cm}{•Managing cruise-ships has been a controversial issue in Australia with an investigation still ongoing into the Ruby Princess. Integrated communication and improved risk assessment are necessary in the future.}} \\
    \multicolumn{1}{|p{25.465em}|}{•        Lack of integrated communication between different government departments} &  \\
    \multicolumn{1}{|p{25.465em}|}{•        Flawed risk assessment} &  \\
    \multicolumn{1}{|p{25.465em}|}{Time required to ramp up testing/ healthcare capability: } & \multicolumn{1}{l|}{\multirow{4}[0]{7.5cm}{•Australia took a long time to ramp up testing and healthcare capability. More detailed planning, and local manufacturing would reduce time required and reduce dependency on international supply chains.}} \\
    \multicolumn{1}{|p{25.465em}|}{•        Limited pandemic preparedness in Australia} &  \\
    \multicolumn{1}{|p{25.465em}|}{•        Limited manufacturing resources in Australia} &  \\
    \multicolumn{1}{|p{25.465em}|}{•        Dependency on international supply chains} &  \\
    \multicolumn{1}{|p{25.465em}|}{Delay to implementing international travel controls:} & \multicolumn{1}{l|}{\multirow{4}[0]{7.5cm}{•Most countries delayed implementing travel controls with limited evidence on effectiveness and WHO recommending against them. More research is required, and the current controls should prompt a rethink from the WHO.}} \\
    \multicolumn{1}{|p{25.465em}|}{•        Lack of previous evidence on effectiveness} &  \\
    \multicolumn{1}{|p{25.465em}|}{•        WHO recommendations against travel controls} &  \\
    \multicolumn{1}{|p{25.465em}|}{•        Issues with information transparency in China} &  \\
          &  \\
    \multicolumn{1}{|p{25.465em}|}{Tasmania breakout} & \multicolumn{1}{l|}{\multirow{3}[0]{7.5cm}{•Use of electronic contract tracing measures such as apps like COVID-SAFE, should help to minimise future outbreaks like that experienced in Tasmania.}} \\
    \multicolumn{1}{|p{25.465em}|}{•        Limits on conventional systems for contact tracing.} &  \\
    \multicolumn{1}{|p{25.465em}|}{Massive economic and social impact} &  \\
    \multicolumn{1}{|p{25.465em}|}{•        Significant job losses} & \multicolumn{1}{l|}{\multirow{3}[0]{7.5cm}{•Public health measures in WA and nationally have come at a high economic and social cost. More research is needed on what control measures are the most cost effective in the Australian context.}} \\
    \multicolumn{1}{|p{25.465em}|}{•        Possible Australian recession and/or depression} &  \\
    \multicolumn{1}{|p{25.465em}|}{•        Significant increases in government debt} &  \\
          &  \\
    \bottomrule
    \end{tabular}%
\end{table}%

\clearpage

\clearpage

\clearpage

\clearpage

\section*{Appendix A - Author's Professional Social Media Timeline Extract}

\textbf{20th March 2020: Higher Learning Progress \& Coronavirus - a positive story}

Having been quiet for a while, I am going to be posting more on our work around the world on coronavirus, balancing the highs \& lows.  Here is a positive story which I trust gives inspiration for the challenges ahead. It's always a joy to see the projects we work with at the early stages come to fruition, especially when it is in the area of learning, research and collaboration.  Integrated Energy were engaged by the Dean of Engineering, Computing \& Mathematics at the University of Western Australia, my alma mater.  Our role was to look forward thirty years into the future \& design how learning, research \& industry collaboration could happen, \& hence shape the concept for EZONE, a once in a generation investment in new facilities for the faculty.  We considered people, process, physical environment, technology \& the resulting business case.  Feedback from students, industry, the profession was very positive - concept we developed in picture 1 - helped create the building in picture 2, 5 years later.  A key capability we flagged was the opportunity for staff \& students to interact digitally independent of location.  This has become very important in recent months, as UWA transitions online ahead of others.

\textbf{25th March 2020: Coronavirus Response at City Level - a positive story}

So I committed to post some more about Integrated Energy actions in coronavirus response.  Today we ran a remote workshop for a City client with 100,000 constituents, we shared on coronavirus global, national and state status and learning, with our review of their existing risk and business continuity plans in light of coronavirus.  We ran them through a range of custom designed scenarios interactively, which helped exercise their Incident Response Team, test their preparations, as well as inform their Executive Leadership Team and Management Leadership team.  From this they saw both risk and opportunity.  Although it was at times emotionally challenging, this knowledge provided them with confidence to make practical decisions going forward, a positive result which helped make our community safer and its future brighter.  For us this is an extension of our practice in these areas for the last thirty years.  Thank you also to our globally connected team for support in this.

\textbf{26th March 2020: Community, Company Transformation \& Coronavirus}

In the spirit of supporting positive action in these challenging times, I thought I would share a little on the evolution of our company operating model and what we offer to our clients, whether they are communities or companies.  From inception 11 years ago, we have always used a lean, agile, globally connected model, whether working in client offices or back at home, so working remotely in an integrated way is built in for us, as well a key area of our client practice. We noted the inception of the virus \& made an offer to government early.  We reviewed our own risks \& opportunities \& updated our plans.  We put in place surveillance of local, national \& global trends \& best practice, leveraging our global network. Added capabilities accordingly.  We upgraded our collaboration \& knowledge sharing systems \& processes.  We adopted further measures of working from home as base model.  We proposed \& demonstrated a revised service offering for our clients, including full digital / remote service delivery.  We successfully deployed this in a HSE critical large scale engagement, shared good practice.  

\textbf{28th March 2020: Positive signs in WA }

The data from Australia, \& Western Australia in particular, is cause for encouragement. While we have natural advantages including distance and low population density, using aligned \& responsive governmental decision making, a well-prepared health system, \& strong public response, is showing in declining rates of new cases, with low mortality thus far. Local government, who we are working with, have a key role going forward in keeping the community connected. With emergency services \& resources sector used to a remote \& integrated operations model, key industries are still online.

\textbf{30th March 2020: Believe in you}
It's been a long time since I've pulled on a medical coat, helped out in an Accident \& Emergency ward, set up a field medical post, trained \& led a team of medics, manned an State Emergency Services evacuation facility, or been on the end of a firehose in an Emergency Services team.   But I remember all of the people I worked with, training at Uni or in the Hospital, in the Regiment, the SES, and in Incident Management teams in twenty countries around the world.  I remember the feeling when an exercise shifts to a "no duff", the real thing. Now is the time when you are our heroes, as my daughter wrote in her drawing - be strong... we believe in you. The data trends show the story of low growth in covid mortality in Australia, benchmarked globally, with many advantages on our side, \& some of the measures in place.  Keep it going... But how can we help from here, to other places in the world, where challenges abound?  Sharing knowledge is one thing, evaluating preparedness \& running realistic team scenario exercises, backed by modelling and analysis is another.  Build knowledge to drive out fear and enable action, for a team, a company, a city.  I know we can do this effectively from here, to anywhere, we can make it work in your language.
\clearpage
\textbf{3rd April 2020: Coronavirus Response – Sharing Good Practice} 

With the harshest of challenges, learning and sharing knowledge becomes even more valuable. A privilege to be invited to share good practice in COVID response at city level, from a West Australian perspective, to a national and international audience earlier in the week. Thank you Dr Hafiz Lateef for the opportunity, and to our clients for allowing this to be shared. I would like to acknowledge the wider Integrated Energy team who contributed, including modelling by Prof Michael Small, medical advisors now including Dr Andrew Clark and Dr Diane Mohen, Dr Andrew Marsh, Jason Scuderi, Cameron Ferstat, our global network, and an example from our friends at Aspen Medical. Please reach out for support – while strong challenges remain everywhere, clear learning from progress here. 

\textbf{8th April 2020: COVID Progress to share - Effective suppression in Australia - but don't let up now }

\begin{figure}[!ht]
\centering
\includegraphics[width=0.7\textwidth]{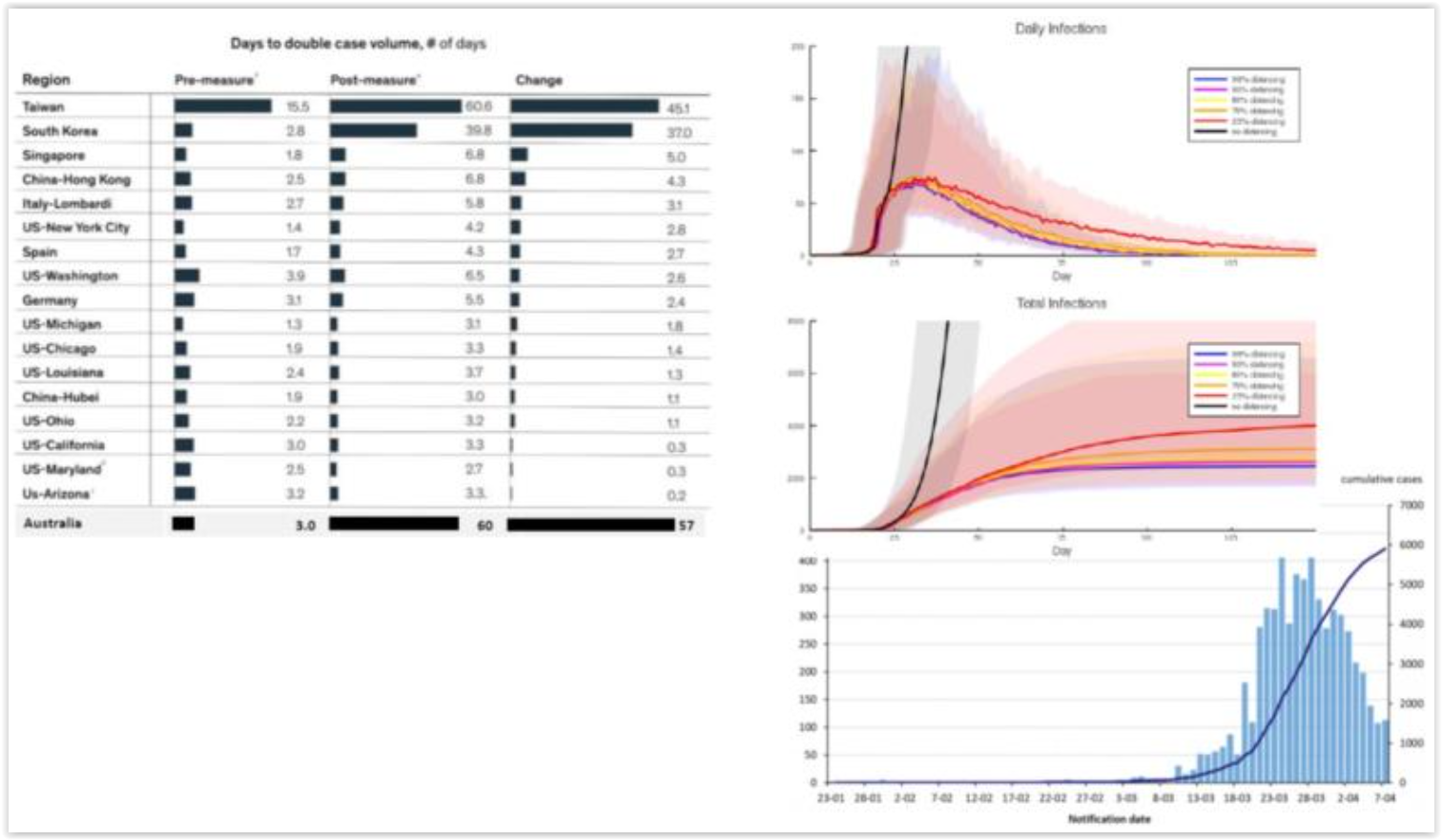}
\caption{Modelling of different adherence to social distancing measures and spread of COVID-19 globally}
\label{fig:Figure 12.}
\end{figure}

As part of client COVID response and recovery planning debrief today - I plotted a telling statistic on COVID suppression. The plot shows the effectiveness of response slowing the spread of COVID, from a range of countries (Figure \ref{fig:Figure 12.}). However, our modelling for Western Australia shows how quickly the epidemic can flare up if social distancing is not maintained. Especially important at Easter. 

\textbf{9th April 2020: From Highway to Hell to Back in Black}

Thoughts go out to those around the world who must feel like they are on the highway to hell right now, working in very difficult conditions.  So I want to send a message of practical benefit, at this time, on this day, which can help those in the middle of this, help others avoid the worst and for us all to recover faster.  The album Back in Black shows that from tragedy can be forged something powerful, enduring, and inspiring.  In working with our clients we have developed a framework that enables them to be prepared for the worst, but also light the way to the opportunities of a brighter future.  We have a document aimed at those working at the local level, which we have built from in our work in COVID response with cities, with success, which allows responses to be benchmarked against a good practice model, and practical measures to link strategy to tactics to action put in place at local and state level, saving both lives and livelihoods.  Please reach out for support wherever you are, so we can build knowledge, drive out fear, through the grief, enable action.  Back in Black.
\clearpage

\textbf{12th April 2020: A Contagion of Hope – Relief \& Recovery for COVID-19 }

\begin{figure}[!ht]
\centering
\includegraphics[width=0.7\textwidth]{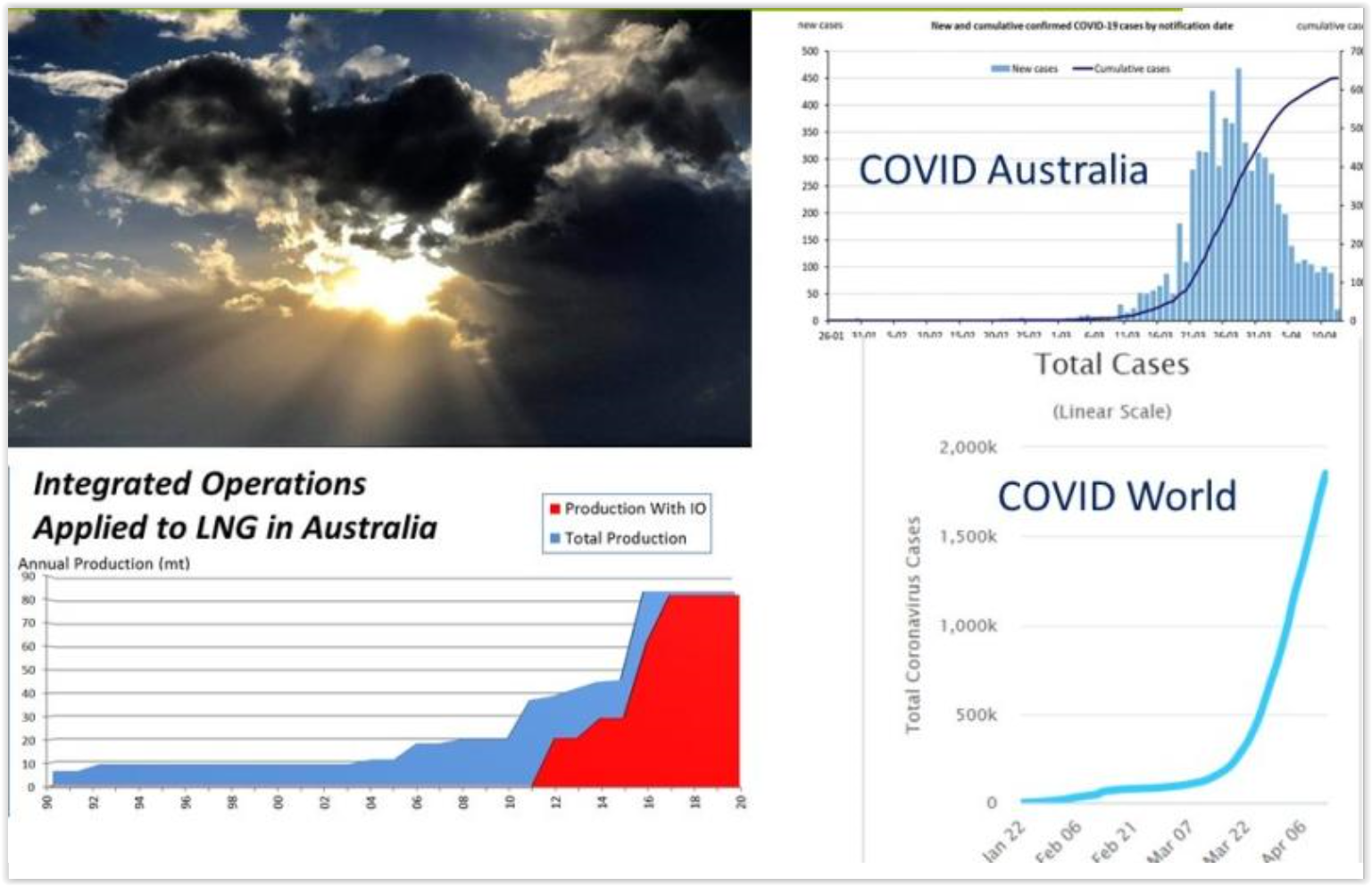}
\caption{Flattening of the COVID-19 Curve in Australia}
\label{fig:Figure 13.}
\end{figure}

Moving into relief \& recovery, across many communities and countries today, decisions are to be made about how to best balance saving lives, \& livelihoods, while keeping people \& society intact. The curve below shows the concept applied in Australian energy - still producing safely. More hope, is the curve of Australia’s COVID cases. In framing relief and recovery, we have a systematic framework which enables the best options \& sequence to be determined in the near term \& longer term, considering your specific values \& circumstances. Do we go back to how things were, or, preserve the best elements of how people combined in response to this challenge, or, is there a healthier, safer, greener, more dynamic \& inclusive model for society which we can build together? In recovery, we can help show how to achieve saving lives, livelihoods and people’s well being at the same time, spreading hope, which shapes behaviour, which leads to success. The corresponding graphic is shown in Figure \ref{fig:Figure 13.}.

\clearpage
\textbf{19th April 2020: Zero Day – The front page of The West tells the story, zero new COVID cases in WA today. }

\begin{figure}[!ht]
\centering
\includegraphics[width=0.7\textwidth]{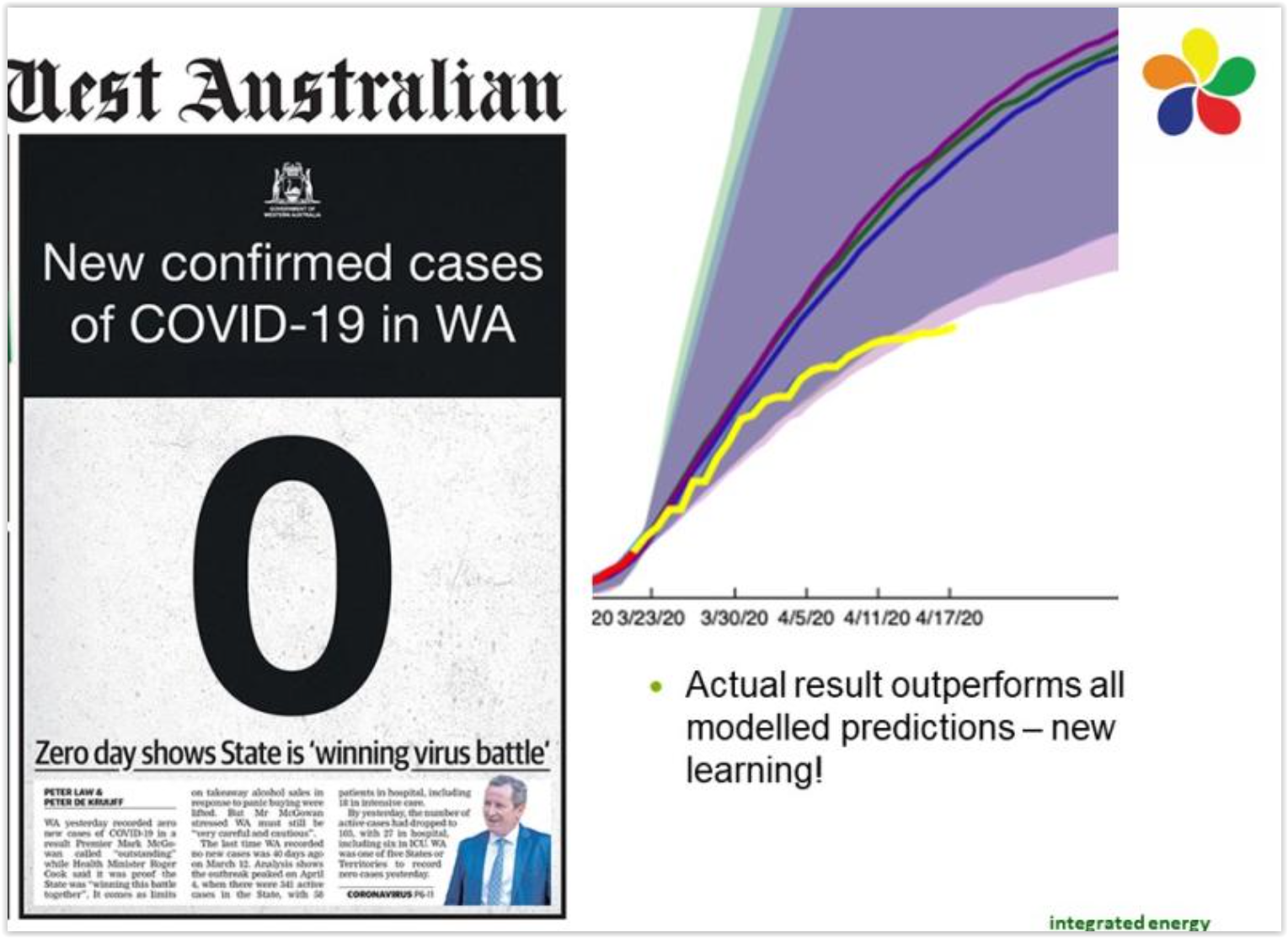}
\caption{First day with no new cases in Western Australia}
\label{fig:Figure 14.}
\end{figure}

We built the models, ran the workshops, mobilised medical services like never before, closed the borders, ramped up Defence, gave the advice … \& people listened. 98+\% compliance to social distancing across the state, the nation. By our analysis the numbers tell the story – most effective response of any state, or close to it. The corresponding graphic is shown in Figure \ref{fig:Figure 14.}.

\textbf{26th April 2020: Mateship, science and COVID response }

\begin{figure}[!ht]
\centering
\includegraphics[width=0.7\textwidth]{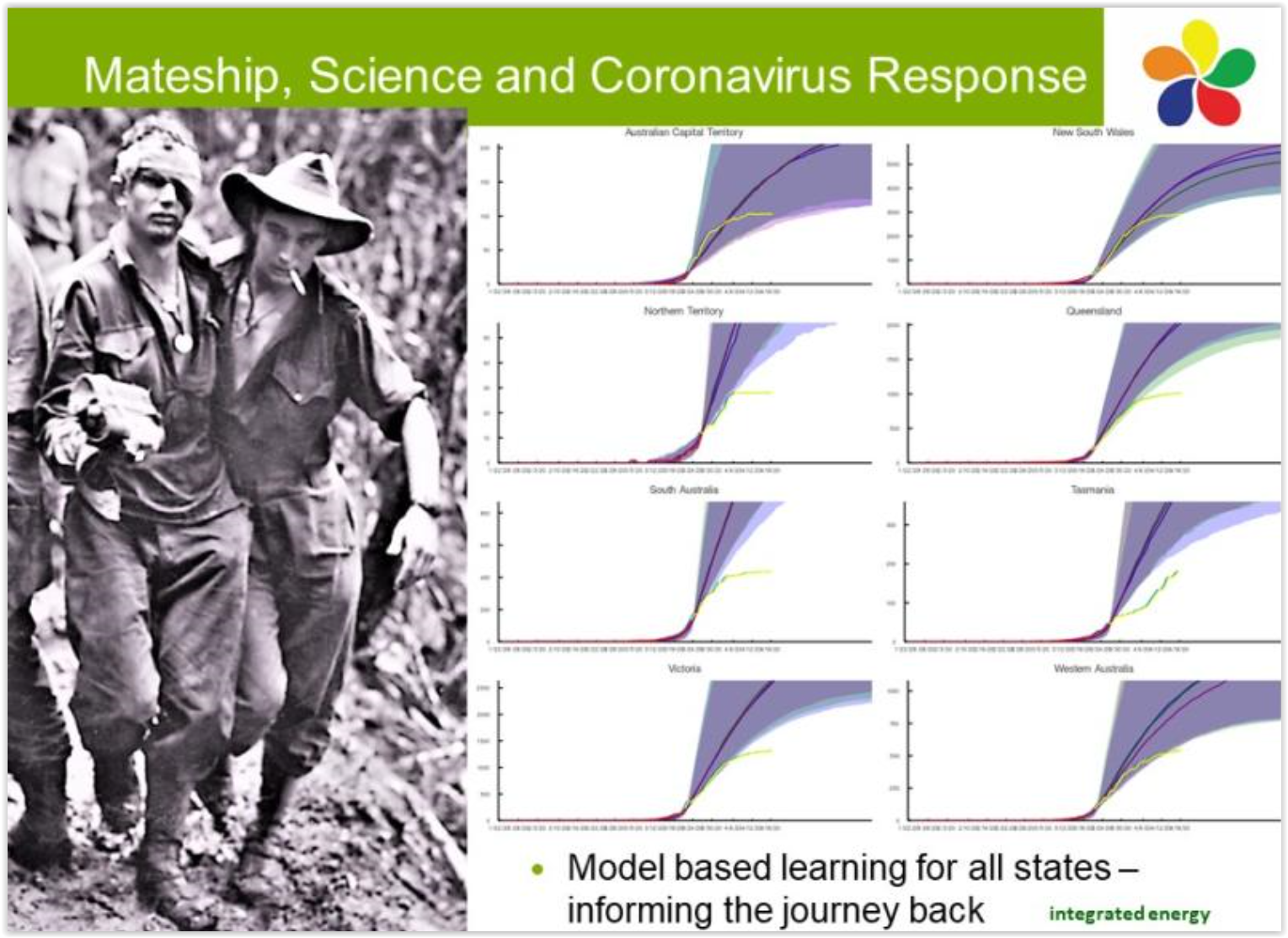}
\caption{Modelled and actual COVID-19 cases in all Western Australian states}
\label{fig:Figure 15.}
\end{figure}

Today I have been comparing the COVID response of all the states in Australia, using our latest model, to identify the critical decisions \& actions which enabled us to be in the good place we are today, with fellow West Australians able to enjoy our ANZAC day holiday in groups of up to 10, for the first time in weeks. The critical period here was clearly from 20--30 March, when every day counted, and numbers were increasingly strongly, 30,000 lives at stake in WA. Before many of the ramped up emergency wards were set up, while testing was still strictly allocated, before the planeloads of PPE could arrive. So what got us through? Ultimately, 2 million + West Australians being prepared to change their behaviour and follow the clear health guidance provided for social distancing. Modify their lives for the good of all. ANZAC spirit. The corresponding graphic is shown in Figure \ref{fig:Figure 15.}.

\textbf{29th April 2020: For the first time today, elimination of COVID from an Australian Territory, ACT. }

Continued focus from the community on hygiene and distancing, readily accessible testing, movement controls, diligent medical care, all lead to the milestone today. We modelled the pandemic for ACT, \& all states, analysed their sequence of events, \& saw their early milestones, so this inspiring result was no surprise. As we begin to lift restrictions, a clear need for practical knowledge transfer to many businesses, so they have a COVID-19 Safety Plan to steer them. 

\textbf{1st May 2020: From Zero Day to your Business Recovery – COVID-19 Safety Plan Training How to safely move your business forward – live guidance }

With a successful COVID-19 response in Australia, we see the need to help thousands of businesses, small and large, safely move your businesses forward into recovery with an effective COVID-19 Safety Plan. A COVID Safety Plan is already a state requirement in Northern Territory to restart, and we recommend it everywhere. We are aiming to convey the “why” as well as the “what” guidelines and give you the information to help develop your Covid safety plan “how”. 

\textbf{3rd May 2020: Undetected cases remain – as we begin a safe journey back}

New modelling shows that although we are seeing zero new cases detected in WA, it is still likely that a significant number of undetected cases remain in our community. Hence authorities care in assessing the timing \& sequence of relaxation measures, \& the need for us all to follow the guidelines, \& download the app. Thanks again Michael Small. The response by West Australians continues to be, in the words of the Premier, world leading. We have distilled learning from WA \& experience nationally in managing COVID-19, compared to global models, \& developed a Pandemic Best Practice Framework which allows us to benchmark pandemic response \& recovery in any organization or govt, rapidly develop a reasonable aspiration, identify gaps \& develop recommendations for improvement. 

\textbf{7th May 2020: The Black Swan of this black swan pandemic – learning, sharing \& planning for safety}
 
As WA, the Black Swan state, achieves a perfect run of zero COVID-19 cases for the last eight days, our efforts turn to learning and sharing the factors that have enabled this success, \& will be necessary to help protect us from a second wave of pandemic in the future. We have formed this into a best practice model we can replicate elsewhere in the world – two plots are shown below for Australia --- firstly reflecting the status today, \& secondly showing the intended status in coming weeks, strengthening our capabilities in sensing, tracking \& tracing, hence allowing current restrictions to be gradually lifted. From analysis of the system dynamics in our pandemic model, \& learning across Australia of rapid track and trace in averting breakouts, the app, if widely applied, can significantly reduce system latency, improve resilience.
Meantime supporting planning for safe business resumption is a focus with refined Business Recovery --- COVID-19 Safety Plan Training:
https://lnkd.in/gKiFjXC
Thanks to our clients in government for support in making this available. A privilege to be invited to brief the Dept of Health on our work.

\textbf{14th May 2020: COVID-19 Safety Plan}
In WA, Our Premier Mark McGowan has just released the guidelines for a COVID-19 Safety Plan:
"COVID Safety Plans are an important part of ensuring that re-opening business does not increase the risk of spreading COIVD-19. Business will need to explain how they will meet conditions and implement the advice set out in these guidelines. A COVID Safety Plan must be completed prior to re-opening. If your business has multiple premises you must prepare a COVID Safety Plan for each premises. You must also display a COVID Safety Plan Certificate in a prominent location." Also, a pleasure to spend an hour this morning with the Assistant Director General of the Department of Health Dr James Williamson, sharing with him the results of our work on COVID-19 response and recovery, and providing advice on COVID measures. Great input from Michael Small. Thank you James for the kind offer for the Department of Health to review our course material.

\textbf{15th May 2020: Kicking Goals for COVID-19 Safety: A Plan for You and Power of the App}
 
Congratulations to Lori Jacobs for being the first to complete the Integrated Energy COVID-19 Safety Plan Course, pass the exam and receive her personalised certificate, on Friday just one day after the Safety Plan requirements were released. Lori found the course “wonderful”, benefitting from the direct dialogue with our lead instructor Dr Andrew Clark. It left her confident with the knowledge to prepare the COVID-19 Safety Plan for her workplace.
https://lnkd.in/gqVPk4c
Meantime our distinctive pandemic model results show the power of the COVIDSafe app, as shown in the figure – with 60\% population signed up the curve reduces strongly from blue to green. Important results from Michael Small recognised in our briefings to WA Dept of Health, WA Chief Scientist \& now shaping thinking from Canberra to the Wall Street Journal.Two days to prepare your mandatory COVID-19 Safety Plan if you are opening up on Monday 18th May.

\textbf{20th May 2020: Wall Street Journal COVID-19 Safety sharing across the world:}

Good to see WSJ catching on to the modelling results from Prof Michael Small \& his work with Integrated Energy in their recent article on the role of managing superspreader events in curbing the COVID-19 pandemic. One figure used to show the impact of measures such as limiting mass gatherings.  And likewise some support from the WSJ forum for my contribution from a West Australian perspective.
https://lnkd.in/dQBGe58
Meantime this morning Michael and I sharing the results of our paper to an audience of 40 Melbourne centric Mathematical and Computational Biology seminar, following a similar session for University of Queensland folks two days ago. And great feedback from attendees at our course on how to develop a COVID-19 Safety Plan for your business, with more scheduled for coming week. Thanks to City Councils in Melville, Victoria Park, and more.

\end{document}